\documentclass[showpacs,preprintnumbers,amsmath,amssymb]{revtex4}
\usepackage{float}
\usepackage{amsmath,amssymb,graphics,epsfig,subfigure}
\usepackage{color}

\usepackage[colorlinks,linkcolor=red,anchorcolor=blue,citecolor=green]{hyperref}
\hypersetup{
    colorlinks=true,
    linkcolor=red,
    filecolor=gray,
    urlcolor=blue,
    citecolor=blue,
}

\begin{document}
\renewcommand{\baselinestretch}{1.3}

\title{Extracting energy via magnetic reconnection from Kerr-de Sitter black holes}

\author{Chao-Hui Wang$^{a,b}$\footnote{ch$\_$W187@126.com}, Cheng-Qun Pang$^{a,b}$\footnote{xuehua45@163.com},
Shao-Wen Wei$^{a,c,d}$ \footnote{weishw@lzu.edu.cn, corresponding author}}

\affiliation{
$^{a}$Joint Research Center for Physics,
Lanzhou University and Qinghai Normal University,
Xining 810000,  People's Republic of China.\\
$^{b}$College of Physics and Electronic Information Engineering, Qinghai Normal University, Xining 810000,  People's Republic of China.\\
$^{c}$Lanzhou Center for Theoretical Physics, Key Laboratory of Theoretical Physics of Gansu Province, School of Physical Science and Technology, Lanzhou University, Lanzhou 730000, People's Republic of China.\\
$^{d}$Institute of Theoretical Physics $\&$ Research Center of Gravitation,
Lanzhou University, Lanzhou 730000, People's Republic of China.}

\begin{abstract}
It has been recently shown that magnetic reconnection can provide us a novel mechanism to extract black hole rotational energy from a Kerr black holes. In this paper, we study the energy extraction from the Kerr-de Sitter black hole via this magnetic reconnection process. The result shows that, with the increase of the cosmological constant, a slowly spinning Kerr-de Sitter black hole can implement the energy extraction better than its Kerr counterpart. Interestingly, although the numerical results show that the maximum values of the power and efficiency slightly decrease with the cosmological constant, Kerr-de Sitter black hole still has significant advantages when the black hole spin is larger than 1 and the dominant reconnection $X$-point is far away from the event horizon. This is mainly attributed to the higher upper spin bound and wider ergosphere in the presence of the cosmological constant. These results uncover the significant effects of the cosmological constant on the energy extraction via the magnetic reconnection process.
\end{abstract}



\maketitle

\section{Introduction}

Since the prediction of general relativity, black holes have played a key role in modern physics. In particular, the recent observations of the black hole binary mergers and images opened a new window for the black hole physics. More intriguing properties about the nature of the gravity and black hole are remained to be disclosed.

As a compact object possessing extreme gravity, a black hole is always related to highly energetic phenomena such as the active Galactic nuclei \cite{McKinney:2004ka,Hawley:2005xs,Komissarov:2007rc,Tchekhovskoy:2011zx} and gamma-ray bursts \cite{Lee:1999se,Tchekhovskoy:2008gq,Komissarov:2009dn}. It is extensively believed that these enormous amounts of energy can originate from the gravitational potential energy, electromagnetic field energy formed in a black hole background, or even more impressively from a black hole itself. Understanding the potential mechanism of energy extraction is helpful to reveal the nature of these observable astrophysical phenomena stimulated by the black holes.

The early study of Christodoulou stated that, for a Kerr black hole characterized by the mass $M$ and spin $a$, a portion of
the black hole mass
\begin{eqnarray}
 M_{\text{irr}}=M\sqrt{\frac{1}{2}\left(1+\sqrt{1-\frac{a^2}{M^2}}\right)},
\end{eqnarray}
related to the black hole horizon entropy $S=4\pi M_{\text{irr}}^2$, is irreducible. Considering the second law of black hole thermodynamics, the maximum amount of energy that can be extracted can reach $E_{\text{rot}}\approx0.29M$ for an extremal spinning Kerr black hole with $a/M$=1. This imposes an upper bound for the extractable rotational energy, which leaves the challenge of how to implement such energy extraction.

For the spinning black hole geometry, there exists an important concept, ergosphere, which corresponds to the asymptotic time translation Killing field $\xi^{a}=(\partial/\partial t)^a$ with
\begin{eqnarray}
\xi^{a}\xi_{a}=g_{tt}>0,
\end{eqnarray}
implying that the tangent vector is space-like. An amazing outcome is that the energy of the test particle can be negative in the ergosphere to a distant observer.

Employing such interesting property, Penrose proposed a first energy extraction mechanism, and now it is known as the Penrose process. The seminal thought experiment of this process is the particle fission, for example, one infalling particle 0 splits into two particles 1 and 2 (i.e. 0$\rightarrow$1+2). Particle 1 is supposed to fall toward the black hole, while particle 2 escapes far away from the black hole. In general, particle 2 carries less energy than the initial particle 0 as expected. The subtlety lies in the fact that such particle fission can occur in the ergosphere of the spinning black hole. Therefore, to a distant observer, particle 1 has a chance to possess a negative energy, which will lead to the result that particle 2 carries more energy than the initial particle 0. Black hole mass and spin will reduce after it absorbs the negative energy particle 1. Therefore, for a distant observer, a net amount of energy is extracted from the black hole, and the black hole slows down. Repeating this process again and again, all the rotational black hole energy will be extracted until the black hole becomes spinless.

Because these two newborn particles should have a very high separation speed and the expected event rate is rare \cite{Bardeen,Wald}, the original Penrose process might be quite limited to extract black hole rotational energy. Nevertheless, it provides us a promising approach. Soon, some improved mechanisms, such as superradiant scattering \cite{Teukolsky}, the particles accelerator \cite{Banados:2009pr,Wei:2010vca}, collisional Penrose process \cite{Piran}, the Blandford-Znajek process \cite{Blandford}, and the magnetohydrodynamic Penrose process \cite{Takahashi} were proposed. Among them, the Blandford-Znajek process taking the advantage of the environmental magnetic field was thought to be the most expected mechanism dominating in the relativistic jets of active Galactic nuclei \cite{McKinney:2004ka,Hawley:2005xs,Komissarov:2007rc,Tchekhovskoy:2011zx} and the gamma-ray bursts \cite{Lee:1999se,Tchekhovskoy:2008gq,Komissarov:2009dn}.

\begin{figure}[htp]
\center{\subfigure[Equatorial plane with $\theta=\frac{\pi}{2}$]{\label{Schematica}
\includegraphics[width=5.6cm]{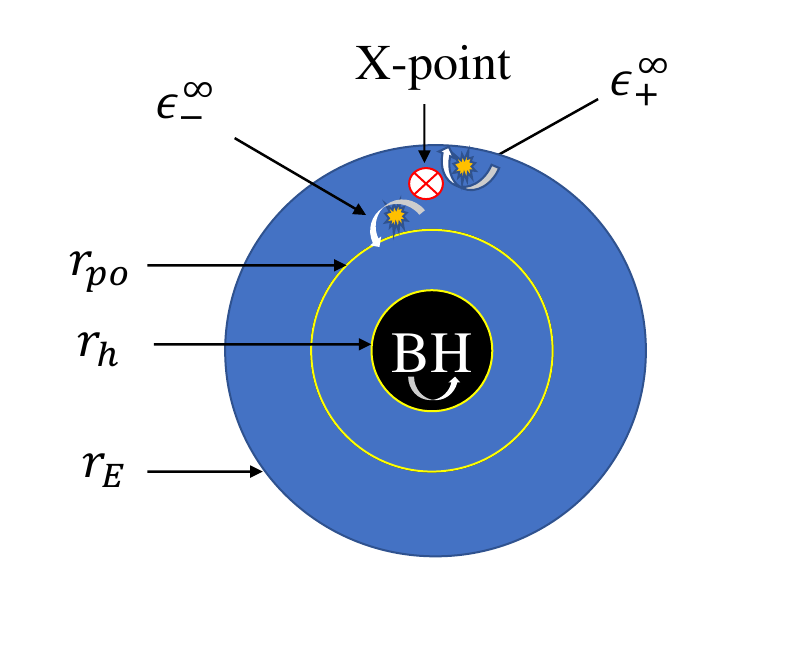}}
\subfigure[Poloidal plane near $\theta=\frac{\pi}{2}$]{\label{Schematicb}
\includegraphics[width=9cm]{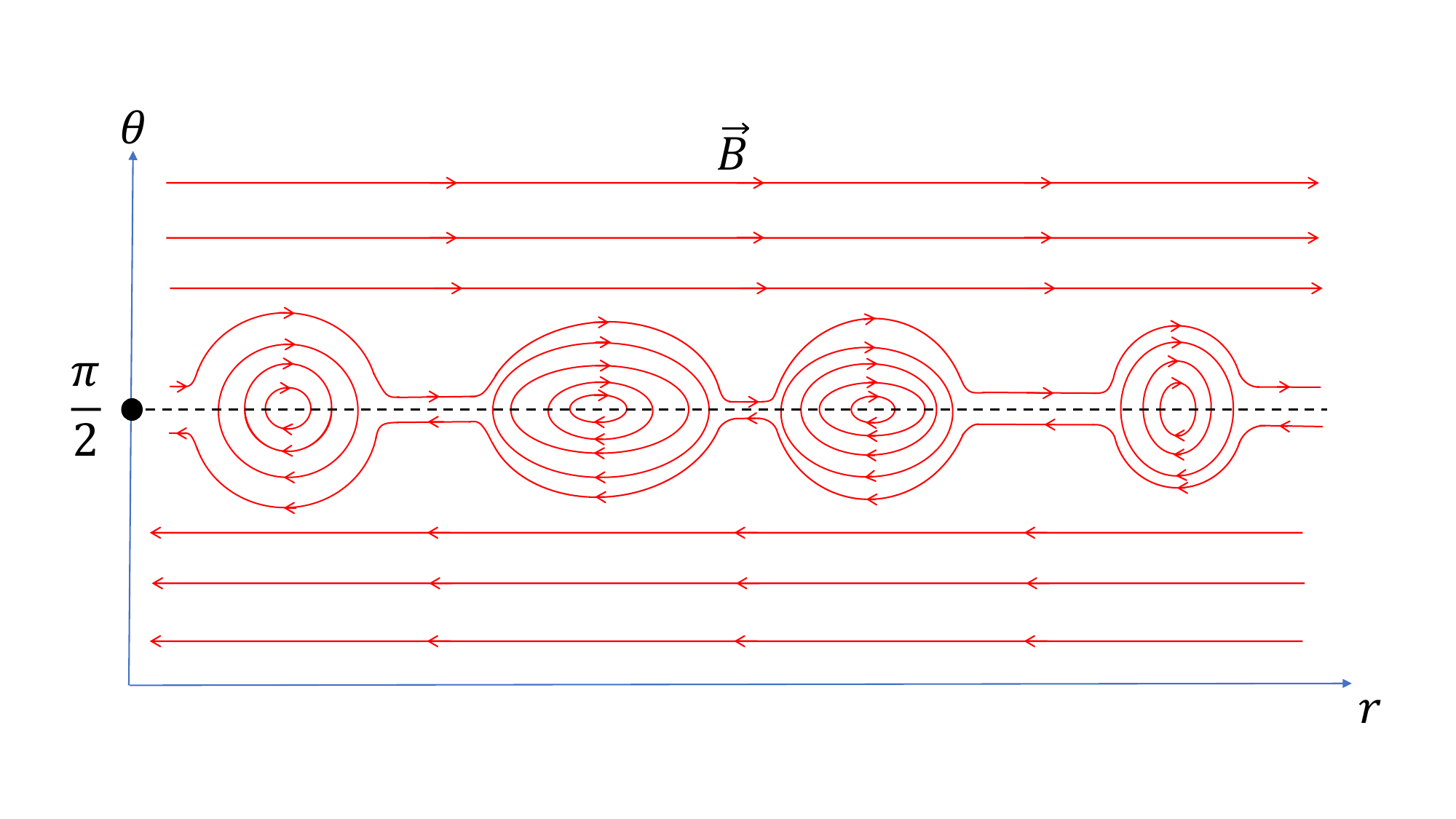}}}
\caption{Schematic illustration of the mechanism of energy extraction from a rotating black hole through the magnetic reconnection within the black hole ergosphere. (a) Features of the black hole on the equatorial plane. $r_{h}$ and $r_{po}$, respectively, denote the radii of the event horizon and circular orbit of the black hole. $r_{E}$ measures the outer boundary of the ergosphere. The ``$X$" point is exactly located inside the ergosphere. The decelerated plasma produced by the magnetic reconnection near that point has negative energy as viewed from infinity, and finally is swallowed by the black hole, while the accelerated plasma with positive energy escapes to infinity. (b) A portion of the antiparallel magnetic field lines on the poloidal plane near the equatorial plane is characterized by the black dashed horizontal line. The antiparallel magnetic field line shrinks and forms the ``$X$" shape points, at which two magnetic reconnection separatrices intersect, on the poloidal plane.} \label{pSchematic}
\end{figure}

Magnetic fields extensively existing near the black holes play an important role in the energy extraction in an astrophysical scenario. In particular, the energy extraction via the magnetic reconnection mechanism proposed in Ref. \cite{Comisso:2020ykg} has attracted much more attention. A schematic illustration of it is given in Fig. \ref{pSchematic}. In Fig. \ref{Schematica}, we list the features of the black hole on the equatorial plane. $r_{h}$ and $r_{po}$, respectively, denote the radii of the event horizon and circular orbit of the black hole, and $r_{E}$ measures the outer boundary of the black hole ergosphere. Near the equatorial plane, it was shown numerically that, for a rapidly rotating black hole, there are antiparallel magnetic field lines on the poloidal plane \cite{Komissarov:2005wj,Koide:2008xr,East:2018ayf,Bransgrove:2021heo,Crinquand:2020reu,Parfrey:2018dnc,Ripperda:2020bpz}. When the aspect ratio of the current sheet exceeds the critical value, the flux ropes/plasmoids will produce the rapid magnetic reconnection process \cite{Comisso:2016pyg,Uzdensky:2014uda,Comisso:2017arh,Sironi} shown in Fig. \ref{Schematicb}. In particular, the positions of the magnetic reconnection are denoted as the $X$-points, at which two magnetic reconnection separatrices intersect \cite{Crinquand:2020reu,L. Sironi:2016,Uzdensky:2010ts,Comisso:2016qvq}. The dominant one has the largest amount of magnetic flux enclosed by the associated separatrix among all these $X$-points. Rapid magnetic reconnection has a certain ability to convert the magnetic energy to the energy of plasma particles, which can escape from the reconnection layer \cite{daughton2009transition,bhattacharjee2009fast}. During such process, one part of the plasma is accelerated and the other part is decelerated. We suppose that the accelerated plasma particles rotate in the same direction as the black hole and escape to infinity, while the decelerated particles move toward the interior of the black hole and are absorbed. Moreover, if this process occurs in the ergosphere described by the blue shaded area in Fig. \ref{Schematica}, the decelerated particles could have negative energy to the distant observer. Then, according to the conservation of energy, the accelerated plasma particles will carry out more energy similar to the Penrose process. When the plasma particles move outside the reconnection region, the magnetic tension relaxes again. Then the magnetic field lines are stretched again due to the frame-dragging effect, and therefore the rapid magnetic reconnection process occurs again. For a rapidly spinning black hole, this process can repeat over and over again stimulated by the dragging effect. To satisfy the requirement that the decelerating particles have negative energy to the observer at infinity, the $X$-point must be located inside the ergosphere of the black hole.

Recently, big progress was achieved by Comisso and Asenjo \cite{Comisso:2020ykg}. The power and efficiency were evaluated for energy extraction via the magnetic reconnection mechanism in the background of a Kerr black hole. The result showed that the magnetic reconnection mechanism is more efficient than the Blandford-Znajek process in some parameter regions. Subsequently, the study was extended to the spinning braneworld black hole \cite{Wei:2022jbi}, non-Kerr black holes \cite{Liu:2022qnr}, with or without broken Lorentz symmetry \cite{Khodadi:2022dff}, and Lorentz breaking Kerr-Sen and Kiselev black holes \cite{Carleo:2022qlv}. All the results indicate that the tidal charge, Lorentz breaking parameter, and other black hole parameters have significant effects on the energy extraction through the reconnection mechanism. Also, such mechanism may dominate in extracting black hole rotational energy.

On the other hand, the positive cosmological constant has attracted a lot of interest, recently, for example, in the expanding Universe \cite{Mottola1985,Dyson2002}, dS/CFT correspondence \cite{Hull1998,Klemm2002,Medved2002,Bousso2002,Ferrero2022,Hikida2022}, and black hole thermodynamics and phase transition \cite{Padmanabhan2002,Kubiznak2017,Dolan2013,Brenna2015}. Different from the asymptotically flat black hole, an extra cosmological horizon appears for the black hole, leading to a different structure of the spacetime. In particular, the cosmological constant modifies the upper bound of the black hole spin such that the rapidly spinning black holes with $a/M>1$ are allowed. Therefore, in this paper, we aim to study the energy extraction via the Comisso-Asenjo process, and examine the effects of the cosmological constant on such process.

This paper is organized as follows. In Sec. \ref{iner}, we give a brief review of the Kerr-dS black hole. In the parameter space, we clearly show the black hole region with or without ergosphere. Several characteristic radii are also examined for different black hole spin and cosmological constant. In Sec. \ref{eevrm}, the effects of the black hole spin and cosmological constant $\Lambda$ on the energy extraction through the magnetic reconnection are investigated. Moreover, the power and efficiency are calculated in Sec. \ref{paevmr}. Finally, the results are summarized and discussed in Sec. \ref{Conclusion}.

\section{Geodesics of a particle around the Kerr-dS black hole}
\label{iner}

In the standard Boyer-Lindquist coordinates $(t,r,\theta,\phi)$ and geometric units $(c=G=1)$, the Kerr-dS geometry with cosmological constant $\Lambda$ is described by the following line element \cite{Carter:1970ea}:
\begin{eqnarray}
 ds^2=g_{tt}dt^2+g_{rr}dr^2+g_{\theta\theta}d\theta^2+g_{\phi\phi}d\phi^2+2g_{t\phi}dtd\phi,\label{mmet}
\end{eqnarray}
where the nonvanishing metric components are
\begin{eqnarray}\label{metric components}
 g_{tt}=-\frac{\Delta_r-\Delta_{\theta}a^2\sin^2\theta}{\rho^2\Sigma^2},\quad
 g_{t\phi}=-\frac{\Delta_{\theta}(r^2+a^2)-\Delta_r }{\rho^2\Sigma^2}a \sin^2\theta,\\
 g_{rr}=\frac{\rho^2}{\Delta_r},\quad
 g_{\theta\theta}=\frac{\rho^2}{\Delta_{\theta}}, \quad
 g_{\phi\phi}=\frac{\Delta_{\theta}(r^2+a^2)^2-\Delta_ra^2 sin^2 \theta}{\rho^2\Sigma^2}\sin^2\theta.
\end{eqnarray}
The metric functions read
\begin{eqnarray}
 \rho^2=r^2+a^2 \cos^2 \theta, \quad
 \Sigma=1+\frac{1}{3}\Lambda a^2, \quad
 \Delta_r=(r^2+a^2)(1-\frac{1}{3}\Lambda r^2)-2Mr, \quad
 \Delta_{\theta}=1+ \frac{1}{3}\Lambda a^2 \cos^2 \theta,
\end{eqnarray}
where $M$ and $a$ are the mass and spin of the black hole.

\begin{figure}[htp]
\center{\includegraphics[width=7cm]{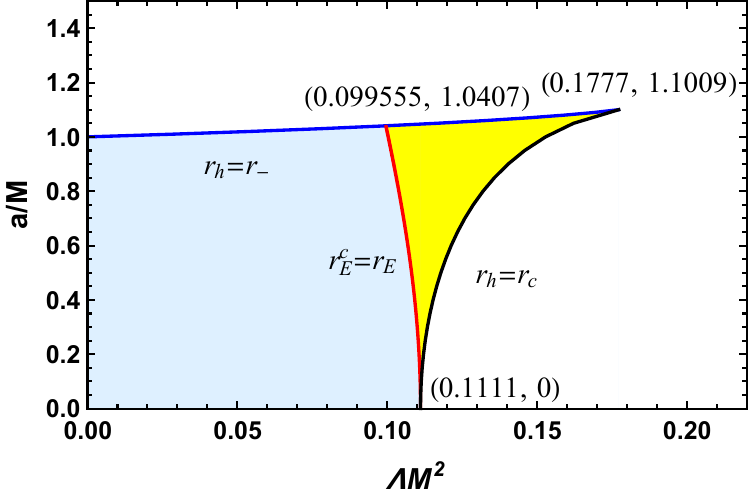}}
\caption{Parameter space of Kerr-dS geometry in the $\Lambda M^2-a/M$ plane. The shaded region in the lower left corner denotes the black hole region with one event horizon at least. The red curve denotes the cases in which the cosmological ergosphere and the black hole ergosphere coincide. On the left side of it marked with light blue color, the black holes possess two ergospheres, while none can be found on the right side.} \label{parameterspace}
\end{figure}

The horizons of the black hole are determined by
\begin{eqnarray} \label{horizons}
 \Delta_r=(r^2+a^2)(1-\frac{1}{3}\Lambda r^2)-2Mr=0.
\end{eqnarray}
This quartic equation of $r$ generally admits four roots, and we denote them as $r_{--}$, $r_-$, $r_h$, and $r_c$ from small to large. The last three ones correspond to the Cauchy horizon, event horizon, and cosmological horizon, respectively. Whereas root $r=r_{--}$ is negative and we abandon it. For the existence of the black hole, one requires $r_{--}< r_-\leq r_h\leq r_c$. The parameter space for Kerr-dS geometry is plotted in Fig. \ref{parameterspace} in the shaded region bounded by the upper blue and lower black solid curves, which, respectively, denote the degeneracies between the black hole event horizon and Cauchy horizon $r_h=r_-$, and the event horizon and the cosmological horizon $r_h=r_c$. Moreover these two curves intersect at the point $(\Lambda M^2,~a/M)\approx (0.1777,~1.1009)$, at which three horizons coincide.

On the other hand, in the ergosphere, one has $g_{tt}>0$. Thus the boundary of the ergosphere can be obtained by solving
\begin{eqnarray} \label{gtt=0}
 g_{tt}=-\frac{\Delta_r-\Delta_{\theta}a^2\sin^2\theta}{\rho^2\Sigma^2}=0.
\end{eqnarray}
In the absence of the cosmological constant $\Lambda$, the Kerr-dS black hole reduces to the Kerr black hole. Via Eq. (\ref{gtt=0}), one can find that the outer boundary of the ergosphere $r=r_E=2M$ in the equatorial plane, which is also independent of the black hole spin. Instead, for a nonzero $\Lambda$, the ergosphere is closely dependent on the parameters $a$. It is quite interesting that by solving (\ref{gtt=0}), we obtain two boundaries at
\begin{eqnarray}
 r^{c}_E&=&\frac{\left(1-i \sqrt{3}\right) \chi_{1}^{2/3}+\sqrt[3]{3} \left(1+i
   \sqrt{3}\right) \Lambda  \left(3-a^2\Lambda \right)}{2\times\sqrt[3]{9\chi_{1} } \Lambda},\\
 r_E&=&\frac{\left(1+i \sqrt{3}\right) \chi_{1}^{2/3}+\sqrt[3]{3} \left(1-i
   \sqrt{3}\right) \Lambda  \left(3-a^2\Lambda \right)}{2\times\sqrt[3]{9\chi_{1} } \Lambda},
\end{eqnarray}
where $\chi_{1}=\sqrt{\Lambda ^3 \left(3 \left(a^2 \Lambda-3\right)^3+729 \Lambda \right)}+27\Lambda^2$. Note that, when $\Lambda\rightarrow0$, $r_E$ recovers that of the Kerr black hole, while $r_{E}^{c}$ tends to infinity. As a result, besides the ergosphere in ($r_h$, $r_E$) near the horizon, an extra one in ($r^{c}_{E}$, $r_{c}$) is present near the cosmological horizon. This one is caused by the cosmological constant, and therefore one can call it the cosmological ergosphere. Despite this fact, in this paper, we only concern ourselves with the energy extraction in the ergosphere near the black hole horizon. A detailed calculation shows that the radii $r^{c}_E$ and $r_E$ coincide at $(3/\Lambda)^{1/3}$, which gives
\begin{eqnarray} \label{rEeqrEcpara}
a^2\Lambda=3-\left(3^5\Lambda\right)^{\frac{1}{3}}.
\end{eqnarray}
We describe the result with the red curve in Fig. \ref{parameterspace}. On the left side of it, there are two ergospheres, while on the right side, no ergosphere is present. Since the energy extraction via magnetic reconnection requires ergospheres, we shall limit our attention to the region with light blue color.

Considering that the magnetic reconnection occurs in the equatorial plane, we are mainly concerned with the equation of motion of the test particle confined in the equatorial plane with $d\theta/d\lambda=0$ and $\theta=\pi/2$. Following the Hamilton-Jacobi equation, the explicit motion of the test particle is given by \cite{Stuchlik:2003dt}
\begin{eqnarray} \label{geodesicequation1}
r^{2} \frac{\mathrm{d} r}{\mathrm{~d} \lambda}=\pm R^{1 / 2}(r), \\
r^{2} \frac{\mathrm{d} \phi}{\mathrm{d} \lambda}=-\Sigma P_{\theta}+\frac{a \Sigma P_{r}}{\Delta_{r}}, \\
r^{2} \frac{\mathrm{d} t}{\mathrm{~d} \lambda}=-a \Sigma P_{\theta}+\frac{\left(r^{2}+a^{2}\right) \Sigma P_{r}}{\Delta_{r}},\label{geoad}
\end{eqnarray}
with
\begin{eqnarray} \label{geodesic equation2}
R(r) &=&P_{r}^{2}-\Delta_{r}\left(m^{2} r^{2}+\Sigma^{2}(a \mathcal{E}-\Phi)^{2} \right), \\
P_{r} &=&\Sigma \mathcal{E}\left(r^{2}+a^{2}\right)-\Sigma a \Phi, \\
P_{\theta} &=&\Sigma(a \mathcal{E}-\Phi),\label{KSigma}
\end{eqnarray}
where $m^2$=1 and 0 for the massive and massless particles. Parameter $\lambda$ is the affine parameter. Energy $\mathcal{E}$ and axial angular momentum $\Phi$ of the test particle are two constants along each geodesic. For the photon with $m^2=0$, it is easy to find that the radii $r_{po}^{+}$ and $r_{po}^{-}$ of the circular corotating and counterrotating photon orbits are given by
\begin{eqnarray}
r_{po}^{+}&=&\frac{6 \left(3-a^2 \Lambda-\sqrt{a^4 \Lambda^2-42 a^2 \Lambda +9} \sin
   \left(\frac{1}{3} \arcsin\chi_{2}\right)\right)}{\left(a^2 \Lambda +3\right)^2},\label{rpho(cor)}\\
r_{po}^{-}&=&\frac{6 \left(3-a^2 \Lambda+\sqrt{a^4 \Lambda ^2-42 a^2 \Lambda +9} \cos
   \left(\frac{1}{3} \arccos\chi_{2}\right)\right)}{\left(a^2 \Lambda \right)^2} \label{rpho(coun)},
\end{eqnarray}
where
\begin{eqnarray}
 \chi_{2}=\frac{1}{3} \left(\frac{1}{a^4 \Lambda ^2-42a^2 \Lambda +9}\right)^{3/2} \left(2a^{10} \Lambda ^4+24 a^8 \Lambda ^3+3 a^6\Lambda ^2 (\Lambda +36)+27 a^4 \Lambda
(11 \Lambda +8)-81 a^2 (11 \Lambda-2)-81\right).
\end{eqnarray}
For the massive particle with $m^2=1$ along timelike geodesics, it can round the black hole at fixed radius. The corresponding Keplerian angular velocity can be obtained by combining with (\ref{geodesicequation1}) and (\ref{geoad}), which gives \cite{Slany:2005vd}
\begin{eqnarray} \label{rph}
\Omega_{K\pm}=\frac{\sqrt{1-\frac{1}{3}\Lambda M^2 r^3}}{a\sqrt{1-\frac{1}{3}\Lambda M^2 r^3}\pm r^{\frac{3}{2}}}.
\end{eqnarray}
The signs ``$\pm$" represent the corotating and counterrotating orbits, respectively.

\begin{figure}[htb]
\center{
\subfigure[ ]{\label{runite}
\includegraphics[width=5.6cm]{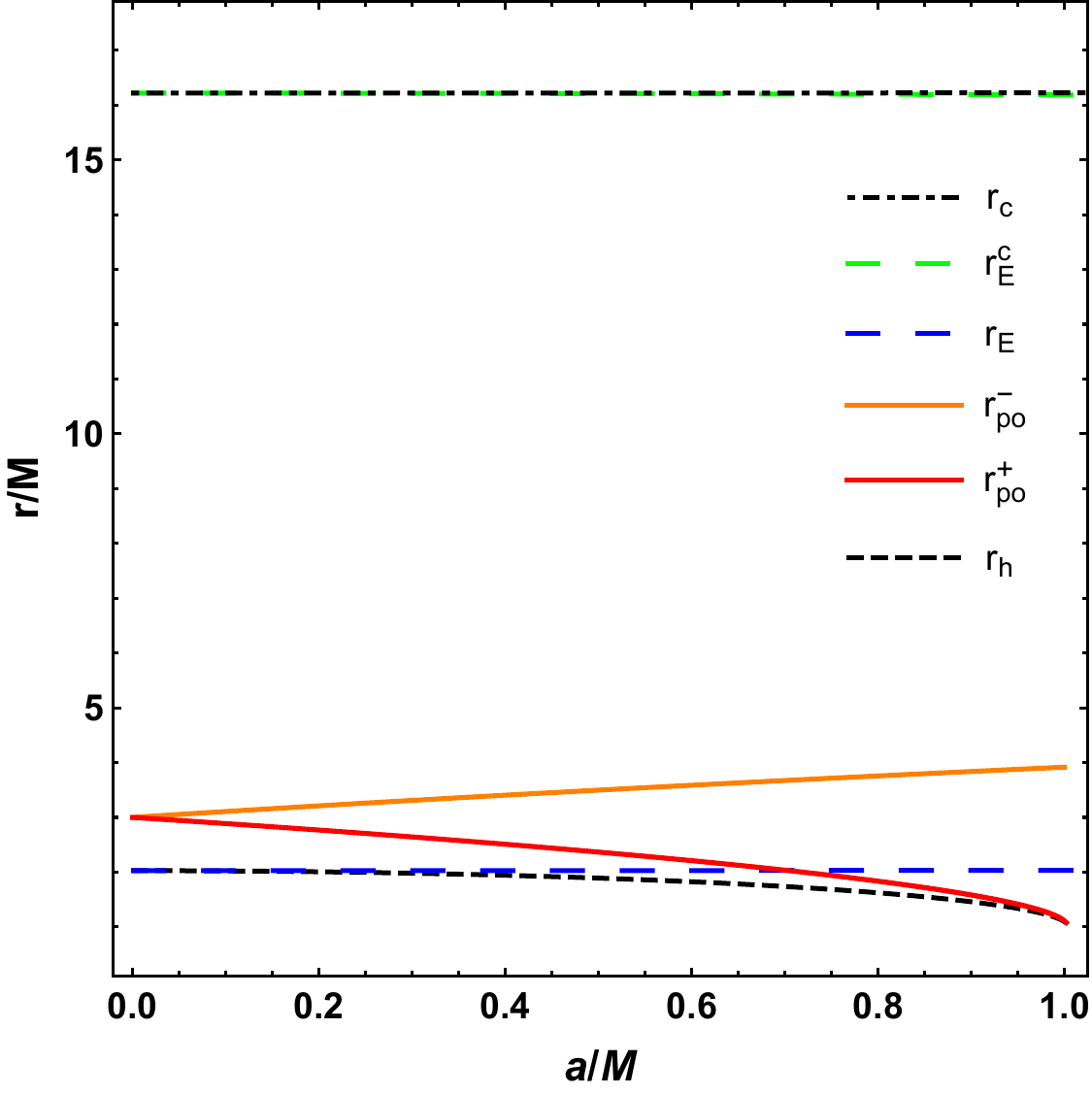}}
\subfigure[ ]{\label{rphre}
\includegraphics[width=5.6cm]{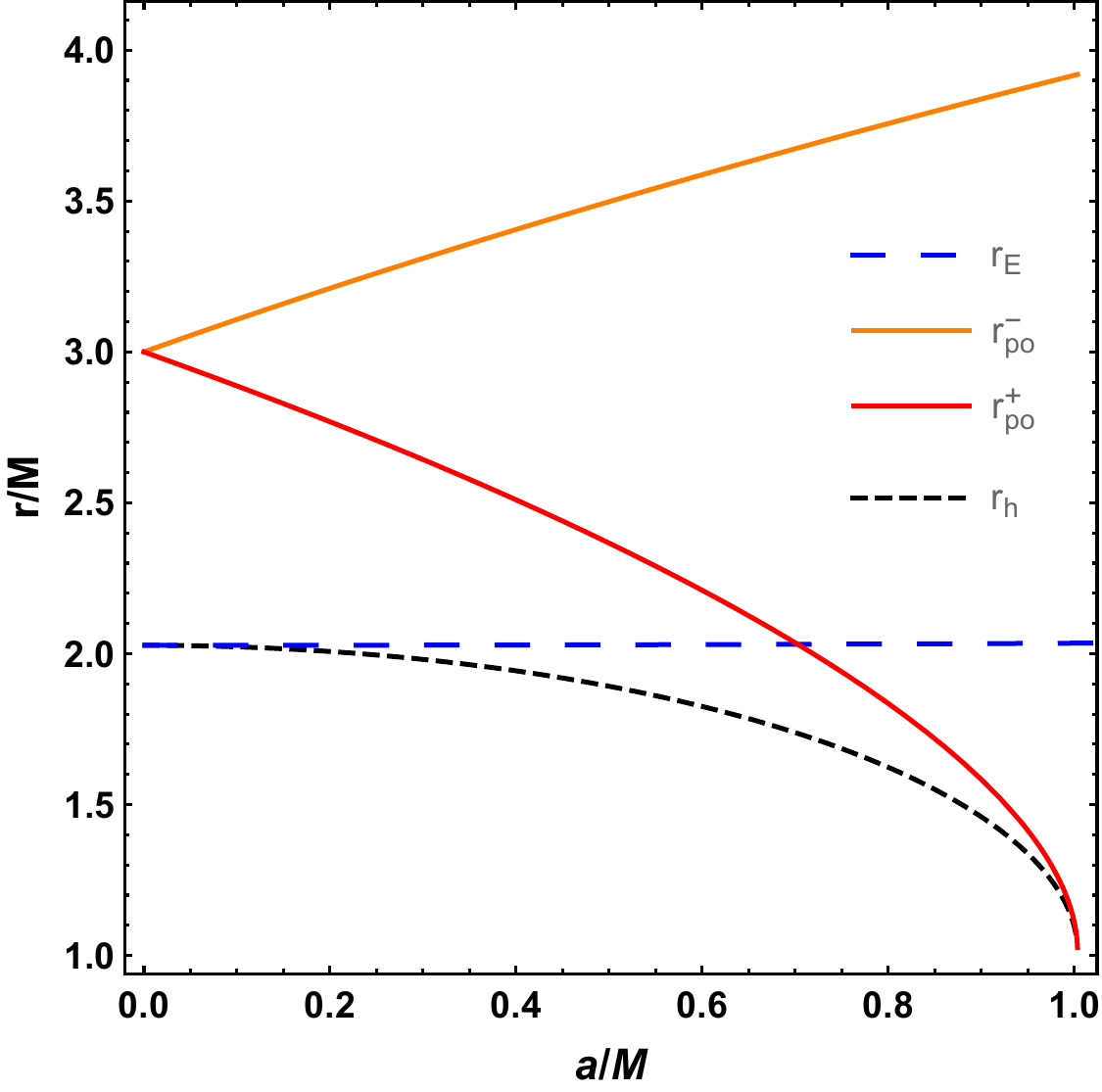}}
\subfigure[ ]{\label{rcre}
\includegraphics[width=5.8cm]{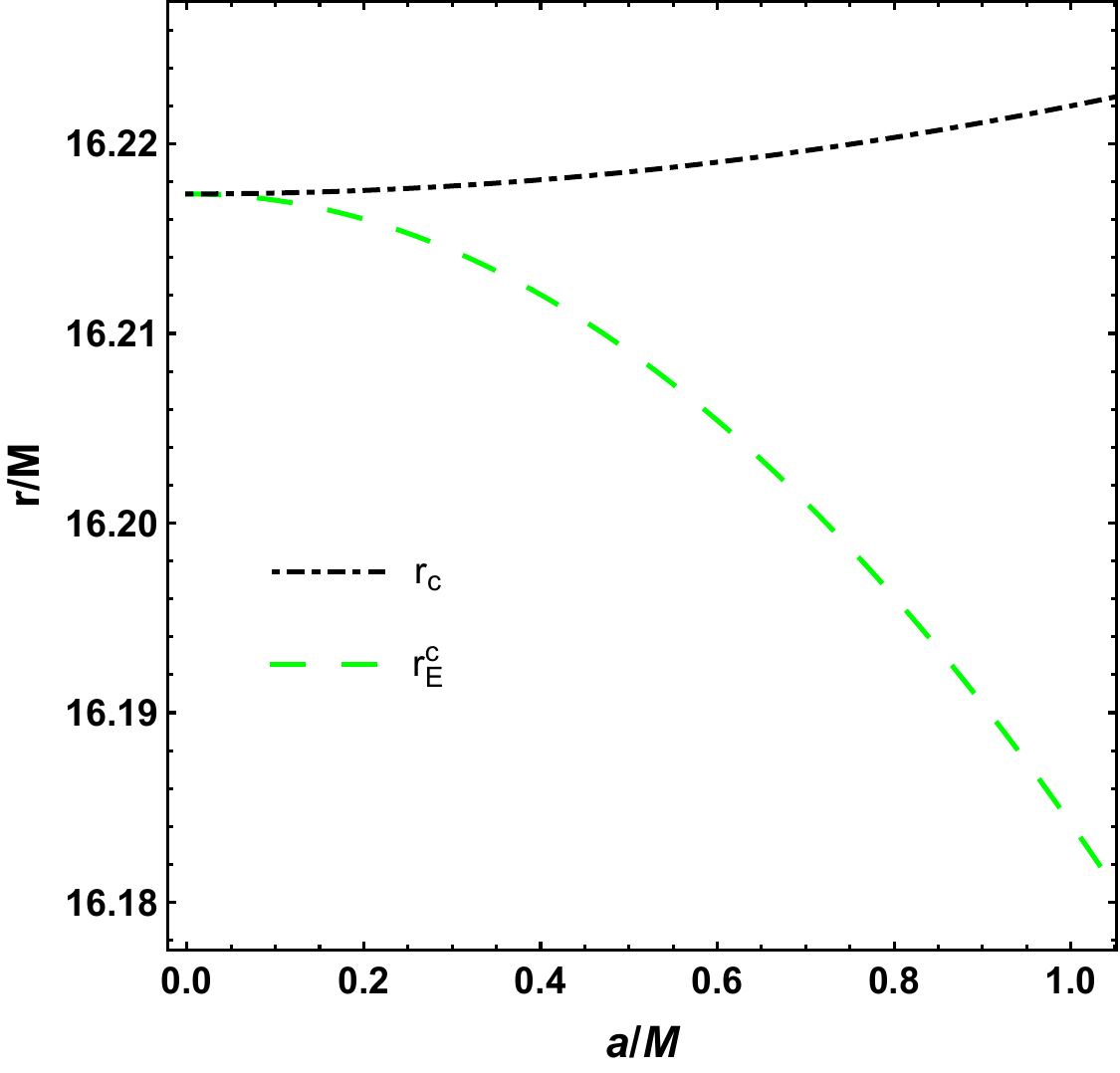}}
\subfigure[ ]{\label{runitelambda_2d}
\includegraphics[width=5.7cm]{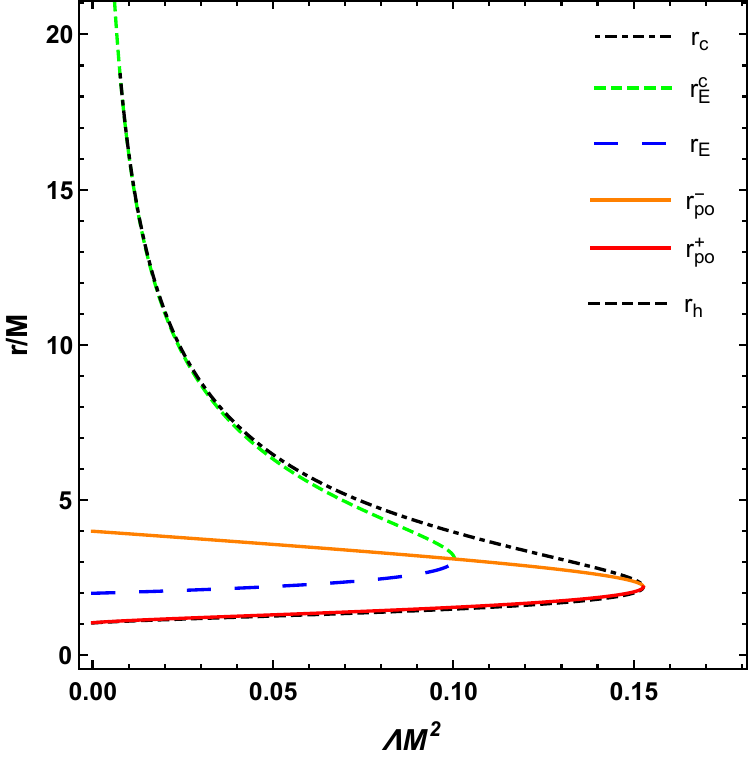}}
\subfigure[ ]{\label{runitelambda_2e}
\includegraphics[width=5.6cm]{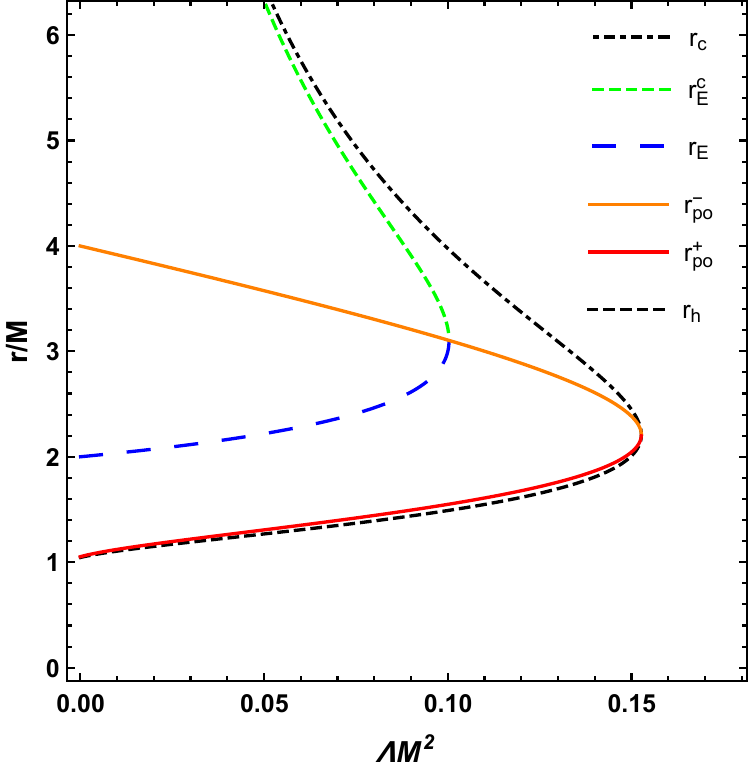}}
\subfigure[ ]{\label{rcrelambda_2f}
\includegraphics[width=5.8cm]{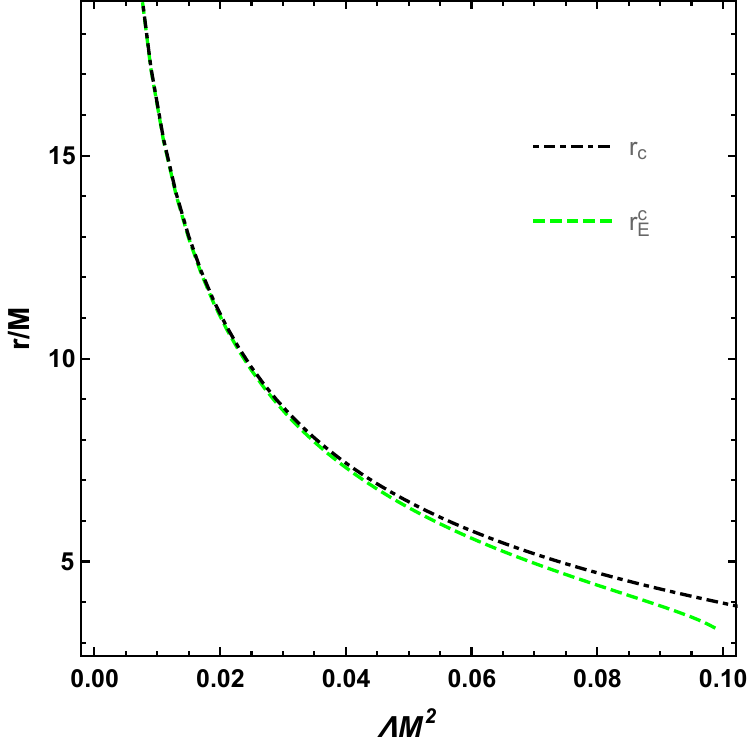}}
}
\caption{The behaviors of the characteristic radii of Kerr-dS black hole. (a)-(c) Variations with the spin parameter $a/M$ for $\Lambda M^2$=0.01. (d)-(f) Variations with the cosmological constant $\Lambda M^2$ for $a/M$ =0.999.}\label{character radill}
\end{figure}

In Fig. \ref{character radill}, we draw the characteristic radii $r_c$, $r_{E}^c$, $r_{E}$, and $r_h$ of the cosmological horizon, outer ergosphere, inner ergosphere, and the event horizon in the backgrounds of Kerr-dS black holes. The radii $r_{po}^+$ and $r_{po}^-$ of the circular corotating and counterrotating photon orbits are also exhibited. In Figs. \ref{runite}-\ref{rcre}, we show these radii as a function of the black hole spin with $\Lambda M^2$=0.01. For $a/M$=0, we find that $r_{E}$ and $r_h$, $r_c$ and $r_{E}^c$, $r_{po}^+$ and $r_{po}^-$ coincide, respectively. With the increase of the black hole spin, they continuously deviate from each other. In particular, the black hole horizon and the circular corotating photon orbit merge at the extremal black hole case with $a/M=1.0033$, which slightly modifies the extremal condition $a/M=1$ of the Kerr black hole. On the other hand, different from the Kerr black hole, there is one extra ergosphere near the cosmological horizon, which is caused by the nonvanishing cosmological constant, acting as a unique property for the Kerr-dS black hole.

Taking $a/M=0.999$, we show these characteristic radii in Figs. \ref{runitelambda_2d}-\ref{rcrelambda_2f}. For this case, the cosmological constant has a maximum value $\Lambda M^2=0.1527$, at which the cosmological horizon and event horizon merge. Obviously, the radii $r_h$ and $r_{po}^{+}$ increase with $\Lambda M^2$, and coincide at its maximum value. It can also be found that $r_{E}^c$ decreases while $r_{E}$ increases with the cosmological constant. Of particular interest is that they meet each other at $\Lambda M^2$=0.1003, smaller than the maximum value $\Lambda M^2=0.1527$. This result indicates that these rotating black holes with $\Lambda M^2\in (0.1003, 0.1527)$ have no ergosphere. As a result, the energy extraction is impossible for such black holes. On the other hand, we can also observe that both the radii $r_c$ and $r_{E}^c$ decreases with $\Lambda M^2$. Their difference is tiny for small $\Lambda M^2$, while it becomes significant when $\Lambda M^2$ approaches its maximum value.

These results show us the generic properties for these characteristic radii, and shall be quite helpful for us to consider the energy extraction via the magnetic reconnection mechanism.

\section{Energy extraction via magnetic reconnection mechanism}
\label{eevrm}

In this section, we will consider the energy extraction via the magnetic reconnection mechanism from the Kerr-dS black hole, and analyze the effects of the cosmological constant on the Comisso-Asenjo process.

In order to analyze the plasma energy mass density conveniently, the locally nonrotating frame, a zero-angular-momentum-observer (ZAMO) frame, is introduced. One advantage of this frame is that in this frame, the spacetime is locally Minkowskian for the observer, and the equations become intuitive. In the ZAMO frame, the line element has the following form:
\begin{eqnarray} \label{zamolineele}
ds^2=\eta_{\mu\nu} d \hat{x}^{\mu}d \hat{x}^{\nu}=-d \hat{t}^2+\Sigma_{i=1}^3 (d \hat{x}^i)^2.
\end{eqnarray}
The transformation of coordinates between these two frames is
\begin{eqnarray} \label{ct1}
d\hat{t}=\alpha dt, \qquad d{\hat{x}}^i=\sqrt{g_{ii}}dx^i-\alpha\beta^i dt,
\end{eqnarray}
where $\alpha=\sqrt{-g_{tt}+g^2_{\phi t}/g_{\phi \phi}}$ and $\beta^i=(0,0,\beta^{\phi})$ are the lapse function and shift vector, respectively. Note that $\beta^{\phi}=\sqrt{g_{\phi \phi}}\omega^{\phi}/\alpha$ with the angular velocity of the frame dragging $\omega^{\phi}=-g_{\phi t}/g_{\phi \phi}$.

As shown above, the ability of the magnetic reconnection to extract energy from black holes is assessed by examining the following two conditions: (i) the formation of negative energy particle to the distant observer and (ii) the escape condition for the plasma particles accelerated/decelerated by the reconnection process in the ergosphere. Taking the one-fluid approximation, the energy-momentum tensor $T^{\mu\nu}$ can be expressed as
\begin{eqnarray} \label{emt}
T^{\mu\nu}=p g_{\mu\nu}+\omega U^{\mu} U^{\nu}+ F^{\mu}_{~\delta}F^{\nu\delta}-\frac{1}{4}g^{\mu\nu}F^{\rho\delta}F_{\rho\delta},
\end{eqnarray}
where $p,~\omega,~U^{\mu},~and~F^{\mu\nu}~$ are the proper plasma pressure,
enthalpy density, four-velocity, and electromagnetic field tensor, respectively. The density of the ``energy at infinity" is calculated as
\begin{eqnarray} \label{density of energy}
e^{\infty}=-\alpha g_{\mu 0}T^{\mu 0}=\alpha \hat{e}+\alpha \beta ^{\phi}\hat{P}^{\phi},
\end{eqnarray}
where the total energy density $\hat{e}$ and the azimuthal component of the momentum density $\hat{P}^{\phi}$ read
\begin{eqnarray} \label{total energy density}
\hat{e}=\omega \hat{\gamma}^2-p+\frac{1}{2}(\hat{B}^2+\hat{E}^2),\\
\hat{P}^{\phi}==\omega \hat{\gamma}^2 \hat{v}^{\phi}+(\hat{B}\times \hat{E})^{\phi}.
\end{eqnarray}
Here $\hat{\gamma}=\hat{U}^0=\sqrt{1-\Sigma_{i=1}^3 (d \hat{v}^i)^2}$ is the Lorentz factor, and $\hat{B}^i=\epsilon^{i j k}\hat{F}_{j k}/2$ and $\hat{E}^i=\eta^{i j}\hat{F}_{j 0}=\hat{F}_{i 0}\hat{v}^{\phi}$ represent the components of the magnetic and electric fields. Meanwhile, $\hat{v}^{\phi}$ is the azimuthal component of the outflow velocity of plasma particles for a ZAMO observer.

The energy at infinity density $e^{\infty}$ can be divided into the hydrodynamic component $e_{hyd}^{\infty}$ and electromagnetic component $e_{em}^{\infty}$,
\begin{eqnarray} \label{total energy density}
e^{\infty}=e_{hyd}^{\infty}+e_{em}^{\infty},
\end{eqnarray}
where
\begin{eqnarray}
e_{hyd}^{\infty}&=&\alpha \hat{e}_{hyd}+\alpha \beta^{\phi} \omega \hat{\gamma}^2\hat{v}^{\phi},\label{ehyd}\\
e_{em}^{\infty}&=&\alpha \hat{e}_{em}+ \alpha \beta^{\phi}(\hat{B}\times \hat{E})_{\phi}. \label{eem}
\end{eqnarray}
Here $\hat{e}_{hyd}=\omega \hat{\gamma}^2-p$ and $\hat{e}_{em}=(\hat{B}^2+ \hat{E}^2)/2$ denote the
hydrodynamic and electromagnetic energy densities observed in the ZAMO frame. Considering an efficient magnetic reconnection process that converts most of the magnetic energy into the kinetic energy of the plasma, the electromagnetic energy at infinity is negligible \cite{Comisso:2020ykg}. Further combining with the approximation of incompressible and adiabatic plasma, the energy density at infinity can be well calculated by
\begin{eqnarray} \label{einfty}
e^{\infty}=e_{hyd}^{\infty}=\alpha \omega \hat{\gamma}(1+\beta^{\phi})-\frac{\alpha p}{\hat{\gamma}}.
\end{eqnarray}
In order to investigate the local reconnection process, it is convenient to introduce a local rest frame $x^{'\mu}=(x^{'0},x^{'1},x^{'2},x^{'3})$ for the bulk plasma with Keplerian angular velocity $\Omega_K$ in the equatorial plane. In this local rest frame, one can choose the directions such that $x^{'1}$ and $x^{'3}$ are parallel to the radial direction $r$ and azimuthal direction $\phi$, respectively. Then, in the ZAMO frame, the corotating Keplerian velocity can be expressed as \cite{Wei:2022jbi}
\begin{eqnarray} \label{vhk}
\hat{v}_{\mathrm{K}} &=\frac{d \hat{x}^{\phi}}{d \hat{x}^{t}}=\frac{d \hat{x}^{\phi} / d \lambda}{d \hat{x}^{t} / d \lambda}=\frac{\sqrt{g_{\phi \phi}} d x^{\phi} / d \lambda-\alpha \beta^{\phi} d x^{t} / d \lambda}{\alpha d x^{t} / d \lambda}
=\frac{\sqrt{g_{\phi \phi}}}{\alpha} \Omega_{\mathrm{K}}-\beta^{\phi}.
\end{eqnarray}
Here we have made use of the coordinate transformation of the vector $\psi$ between the ZAMO frame and Boyer-Lindquist coordinates
\begin{eqnarray} \label{vhk}
\hat{\psi}^0=\alpha \psi^0, \qquad \hat{\psi}^i=\sqrt{g_{ii}\psi^i}-\alpha\beta^i\psi^0.\\
\hat{\psi}_0=\frac{\psi^0}{\alpha}+\sum_{i=1}^3\frac{\beta_i}{\sqrt{g_{ii}}}\psi_i, \qquad \hat{\psi_i}=\frac{\psi_i}{\sqrt{g_{ii}}}.
\end{eqnarray}
Finally, employing the relativistic adiabatic incomprehensible ball approach, the hydrodynamic energy at infinity per enthalpy of plasma is \cite{Comisso:2020ykg}
\begin{eqnarray} \label{epminf}
{\epsilon_{\pm}^{\infty}=\alpha \hat{\gamma}_{K}\left(\left(1+\beta^{\phi} \hat{v}_{\mathrm{K}}\right) \sqrt{1+\sigma_{0}} \pm \cos \xi\left(\beta^{\phi}+\hat{v}_{\mathrm{K}}\right) \sqrt{\sigma_{0}}-\frac{\sqrt{1+\sigma_{0}} \mp \cos \xi \hat{v}_{\mathrm{K}} \sqrt{\sigma_{0}}}{4 \hat{\gamma}_{\mathrm{K}}^{2}\left(1+\sigma_{0}-\cos^{2} \xi \hat{v}_{\mathrm{K}}^{2} \sigma\right)}\right),}
\end{eqnarray}
where $\hat{\gamma}_K=1/\sqrt{1-\hat{v}_K^2}$ is the Lorentz factor corresponding to $\hat{v}_K$, $\sigma_0=B_0^2/\omega_0$ is the plasma magnetization upstream of the reconnection layer, and $\xi$ is the orientation angle between the magnetic field lines and the azimuthal direction in the
equatorial plane.

In order to achieve this energy extraction mechanism, we require that the decelerated plasma has negative energy, while the accelerated plasma has positive energy for the distant observer \cite{Comisso:2020ykg},
\begin{eqnarray}\label{condition}
\epsilon_{-}^{\infty}<0, \quad \Delta \epsilon_{+}^{\infty}=\epsilon_{+}^{\infty}-\left(1-\frac{\Gamma}{\Gamma-1} \frac{p}{w}\right)=\epsilon_{+}^{\infty}>0,
\end{eqnarray}
for a relativistic hot plasma with the polytropic index $\Gamma=4/3$.

\begin{figure}[htb]
\center{
\subfigure[$r/M=1.3,~a/M=0.99,~\xi=\pi/12$]{\label{EPEM1p3_3a}
\includegraphics[width=5.4cm]{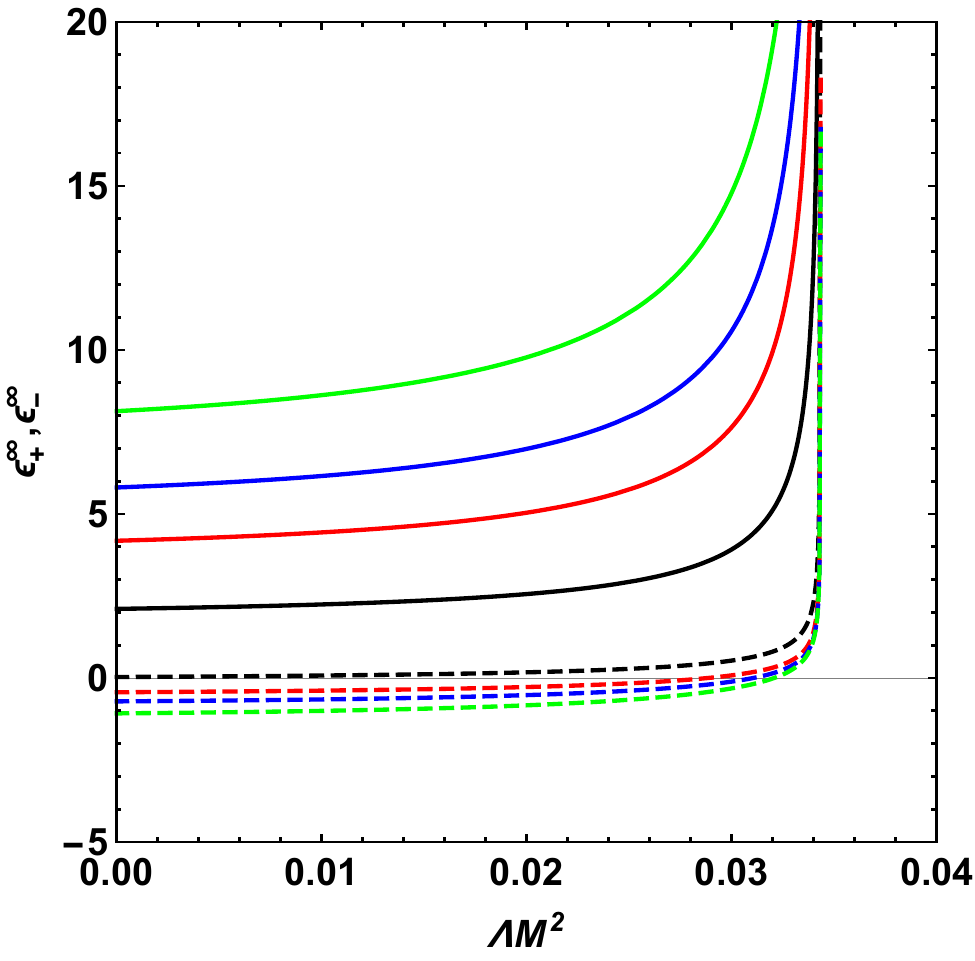}}
\subfigure[$r/M=1.8,~a/M=0.9,~\xi=\pi/12$]{\label{EPEM1p8_3b}
\includegraphics[width=5.5cm]{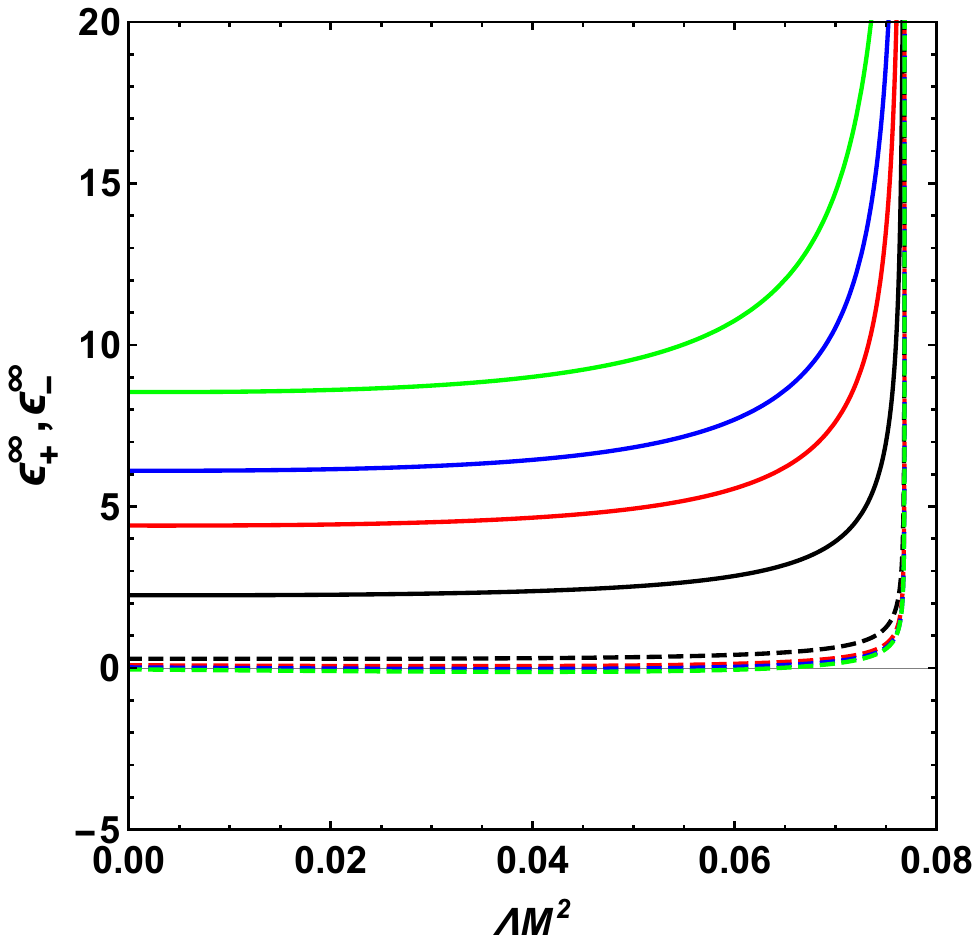}}
\subfigure[$r/M=2,~a/M=0.9,~\xi=\pi/12$]{\label{EPEM2_3c}
\includegraphics[width=5.4cm]{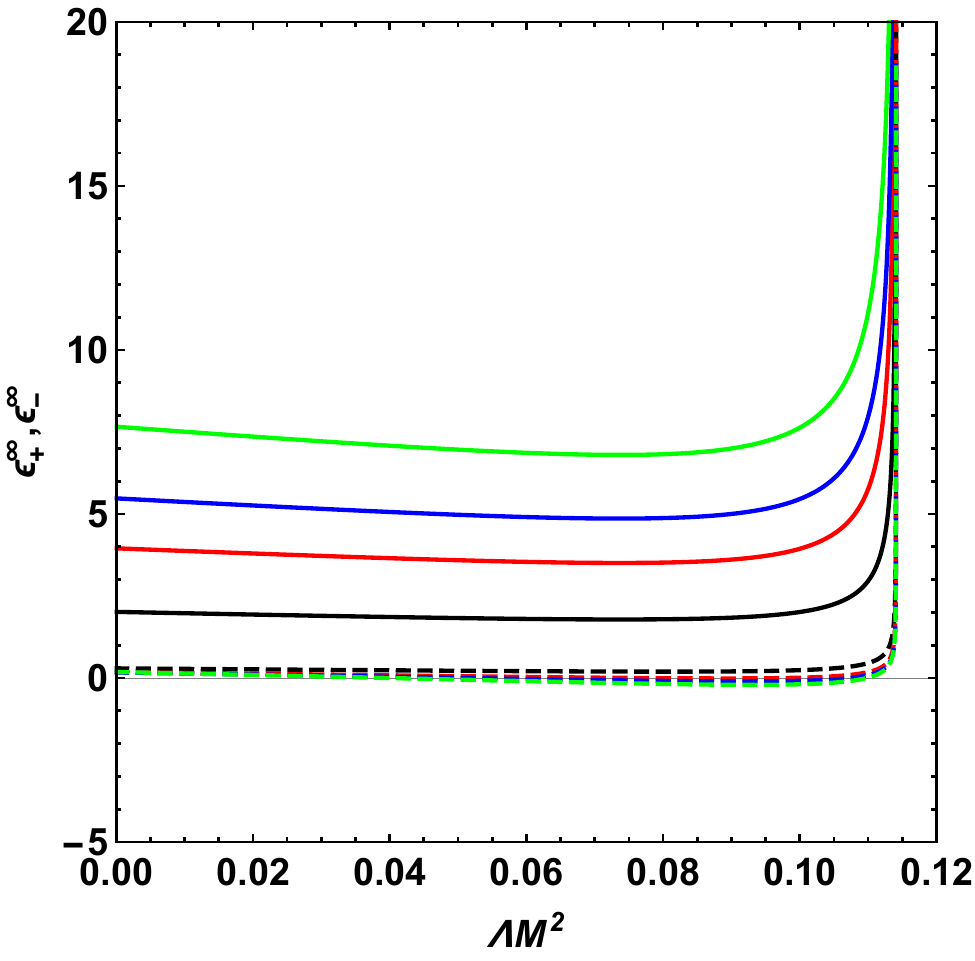}}
\subfigure[$r/M=1.3,~a/M=0.999,~\xi=\pi/12$]{\label{EPEM1p3_3d}
\includegraphics[width=5.5cm]{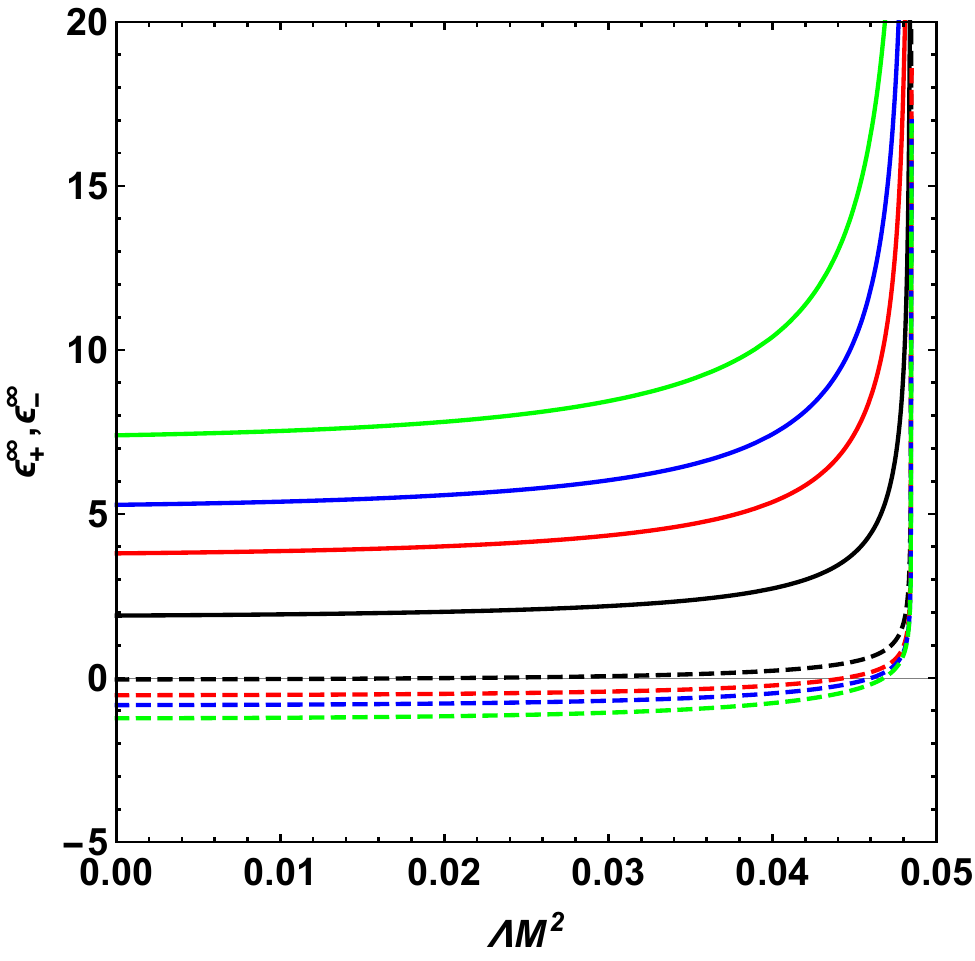}}
\subfigure[$r/M=1.8,~a/M=0.999,~\xi=\pi/12$]{\label{EPEM1p8_3e}
\includegraphics[width=5.5cm]{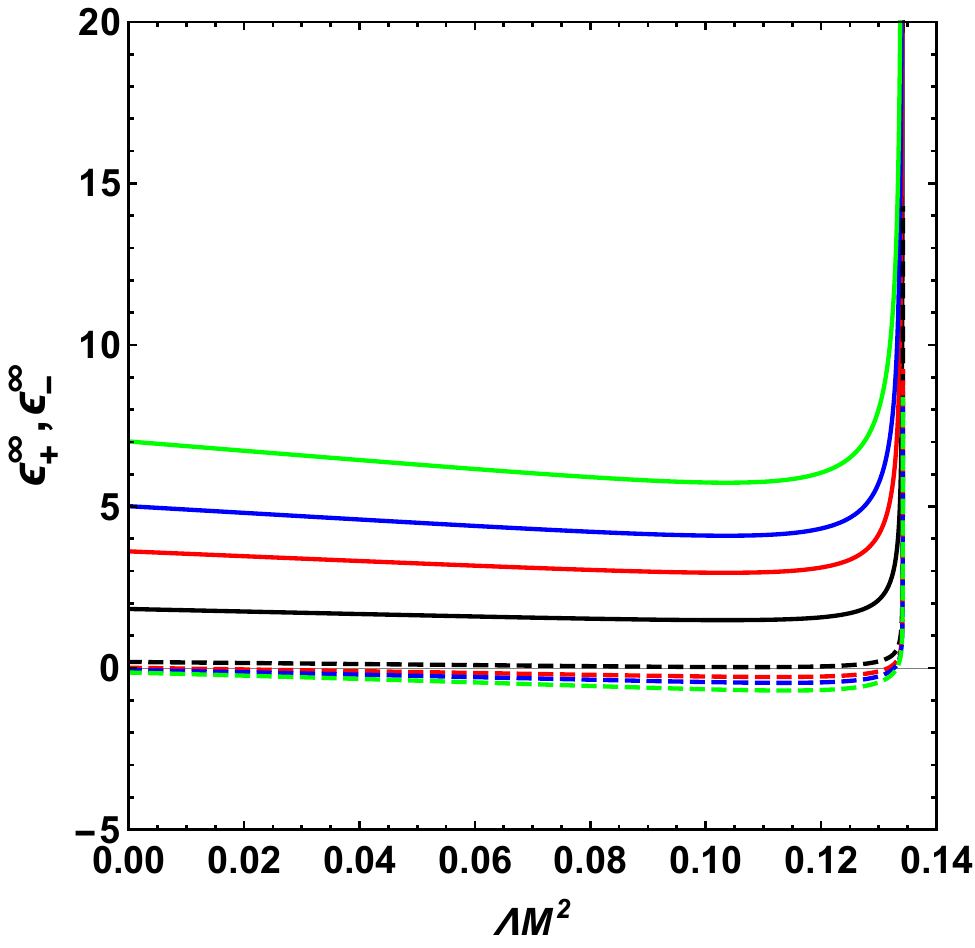}}
\subfigure[$r/M=2,~a/M=0.999,~\xi=\pi/12$]{\label{EPEM2_3f}
\includegraphics[width=5.3cm]{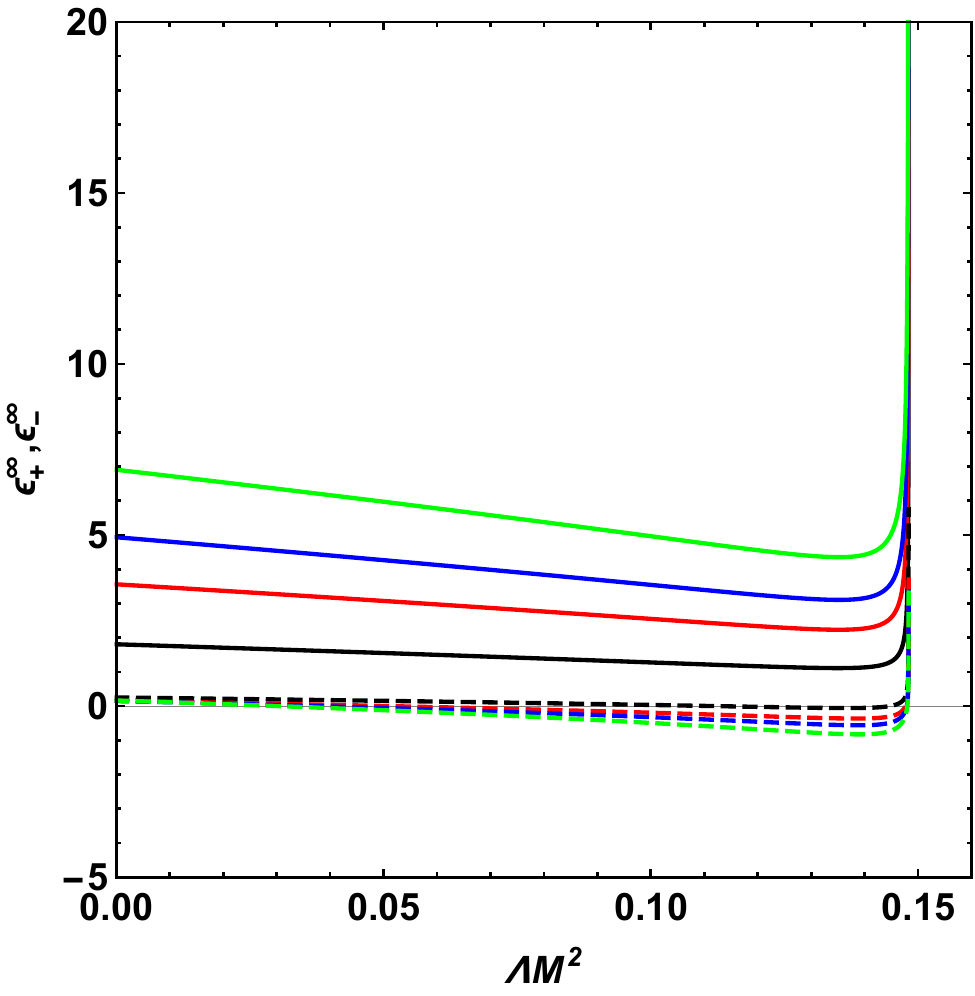}}
}
\caption{The behaviors of $\epsilon_{+}^{\infty}$ (solid curves) and $\epsilon_{-}^{\infty}$ (dashed curves) in terms of the cosmological constant $\Lambda M^2$. The plasma magnetization parameter $\sigma_0$=1, 5, 10, and 20 from bottom to top for solid curves and from top to bottom for dashed curves. }\label{EPEM}
\end{figure}

Now let us examine the behaviors of the energy at infinity per enthalpy $\epsilon_{+}^{\infty}$ and $\epsilon_{-}^{\infty}$ (\ref{epminf}). As has been shown in Ref. \cite{Wei:2022jbi}, the energy extraction from the magnetic reconnection prefers a small orientation angle $\xi$. In order to meet this requirement, we set $\xi=\pi/12$, and then study the effect of the cosmological constant $\Lambda$ on the energy at infinity per enthalpy. Taking certain values of these parameters, we plot $\epsilon_+^{\infty}$ (solid curves) and $\epsilon_-^{\infty}$ (dashed curves) in Fig. \ref{EPEM}. A universal result is that both $\epsilon_+^{\infty}$ and $\epsilon_-^{\infty}$ blow up at some certain values of $\Lambda$. However, these values do not reach the maximum values of $\Lambda$ at which the Kerr-dS black holes exist, see Fig. \ref{parameterspace}. The reason is that for these values of black hole spin and cosmological constant, the locations of the $X$-point meet the corresponding unstable circular photon orbits, at which the energy and angular momentum of the massive particles circling the black hole diverge \cite{Delgado:2021jxd}. Meanwhile, the radius of the unstable circular photon orbit is also a bound of the timelike circular orbit with a Keplerian angular velocity (\ref{rph}).

Another universal result indicated from Fig. \ref{EPEM} is that, with the increase of $\Lambda$, $\epsilon_+^{\infty}$ is always positive, while $\epsilon_-^{\infty}$ can take negative values before it blows up. This result suggests that the energy extraction via the magnetic reconnection mechanism heavily relies on $\epsilon_-^{\infty}$ rather than $\epsilon_+^{\infty}$. So when one focuses on the allowed parameter regions for the energy extraction, condition $\epsilon_-^{\infty}<0$ will be enough. Also, the larger plasma magnetization parameter $\sigma_0$ will produce lower $\epsilon_-^{\infty}$ as expected.

Moreover, it is indicated from Figs. \ref{EPEM1p8_3b} and \ref{EPEM1p8_3e} or Figs. \ref{EPEM2_3c} and \ref{EPEM2_3f} that, for the fixed orientation angle $\xi$ and $X$-point, the region of the cosmological constant $\Lambda$ satisfying $\epsilon_-^{\infty}<0$ enlarges with the black hole spin. Combining with Figs. \ref{EPEM1p3_3d}, \ref{EPEM1p8_3e}, and \ref{EPEM2_3f}, we also find that the range of $\Lambda$ allowed by the magnetic reconnection process becomes wider with the increase of the position of the $X$-point when the black hole spin $a$ and particle orientation angle $\xi$ are fixed.

Another interesting result is presented in Fig. \ref{EPEM2_3f}. For the Kerr black hole with $\Lambda=0$, it is impossible to extract the black hole energy. However, we see that $\epsilon_-^{\infty}$ turns to negative near $\Lambda M^2$=0.05. As a result, in some certain parameter regions, energy extraction from the magnetic reconnection can be implemented for the Kerr-dS black hole rather than its Kerr counterpart with the same spin.

\begin{figure}[htp]
\center{\includegraphics[width=6cm]{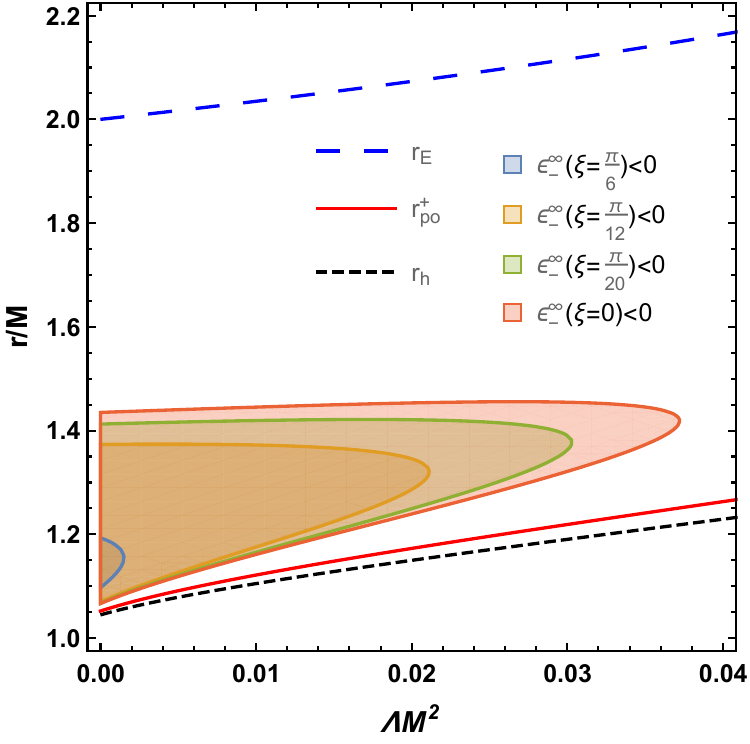}}
\caption{Regions of parameter space $\Lambda M^2-r/M$ for $\epsilon_{+}^{\infty}>0$ and $\epsilon_{-}^{\infty}<0$ with $a/M=0.999,~\sigma_0=1$.} \label{Lambdar_4}
\end{figure}

As shown above, the condition (\ref{condition}) to implement the energy extraction is feasible. Here we would like to consider these possible regions in the parameter space. First, let us focus on the $\Lambda M^2-r/M$ parameter space. From our above study, we see that the condition can be satisfied with negative $\epsilon_-^{\infty}$, so we shall ignore $\epsilon_+^{\infty}$. Taking the black hole spin $a/M$=0.999 and plasma magnetization parameter $\sigma_0$=1 as an example, we exhibit the region of negative $\epsilon_-^{\infty}$ in shaded regions in Fig. \ref{Lambdar_4} with the orientation angle $\xi=0$, $\pi/20$, $\pi/12$, and $\pi/6$. All these shaded regions fall in the ergosphere bounded by these two dashed curves, representing the radii of the ergosphere boundary and black hole horizon. With the decrease of $\xi$, the allowed regions in the $\Lambda M^2-r/M$ parameter space enlarge. The maximum cosmological constant can extend to $\Lambda M^2=0.037$; whereas when $\xi>\pi/6$, the allowed region is quite small and shall be negligible. On the other hand, we can also find that the range of $r$ for the $X$-points decreases with $\Lambda M^2$. This indicates that the presence of the cosmological constant will shrink the region of the energy extraction. However, the cosmological constant raises the maximum spin of the black holes, which could provide us a new chance to study the extraction of energy for the rapidly spinning black hole with $a/M>1$.

\begin{figure}[htp]
\center{
\subfigure[$\xi=\pi/12,~\Lambda M^2=0$]{\label{ar_5a}
\includegraphics[width=5cm]{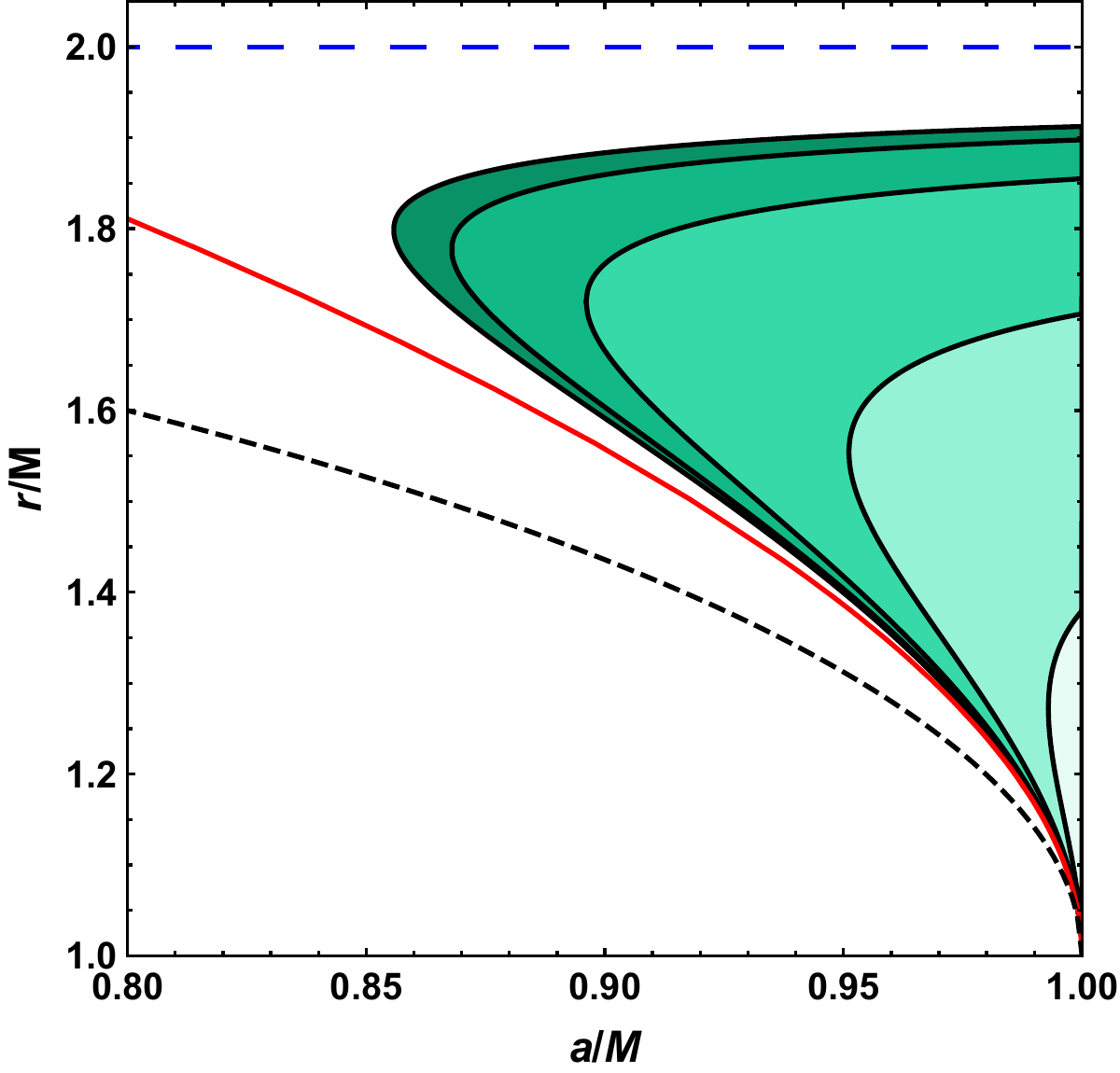}}
\subfigure[$\xi=\pi/12,~\Lambda M^2=0.01$]{\label{ar_5b}
\includegraphics[width=5cm]{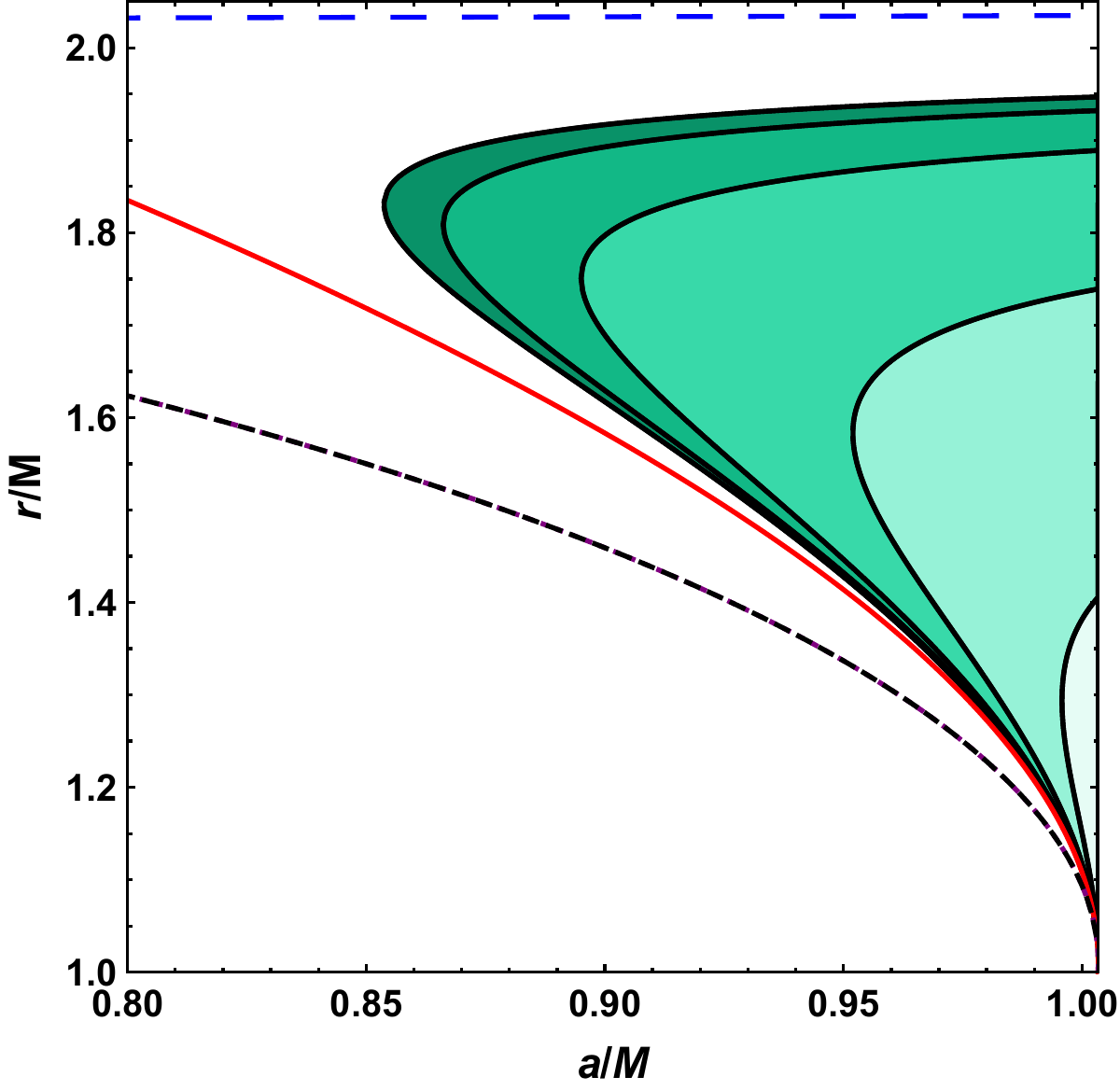}}
\subfigure[$\xi=\pi/12,~\Lambda M^2=0.03$]{\label{ar_5c}
\includegraphics[width=5cm]{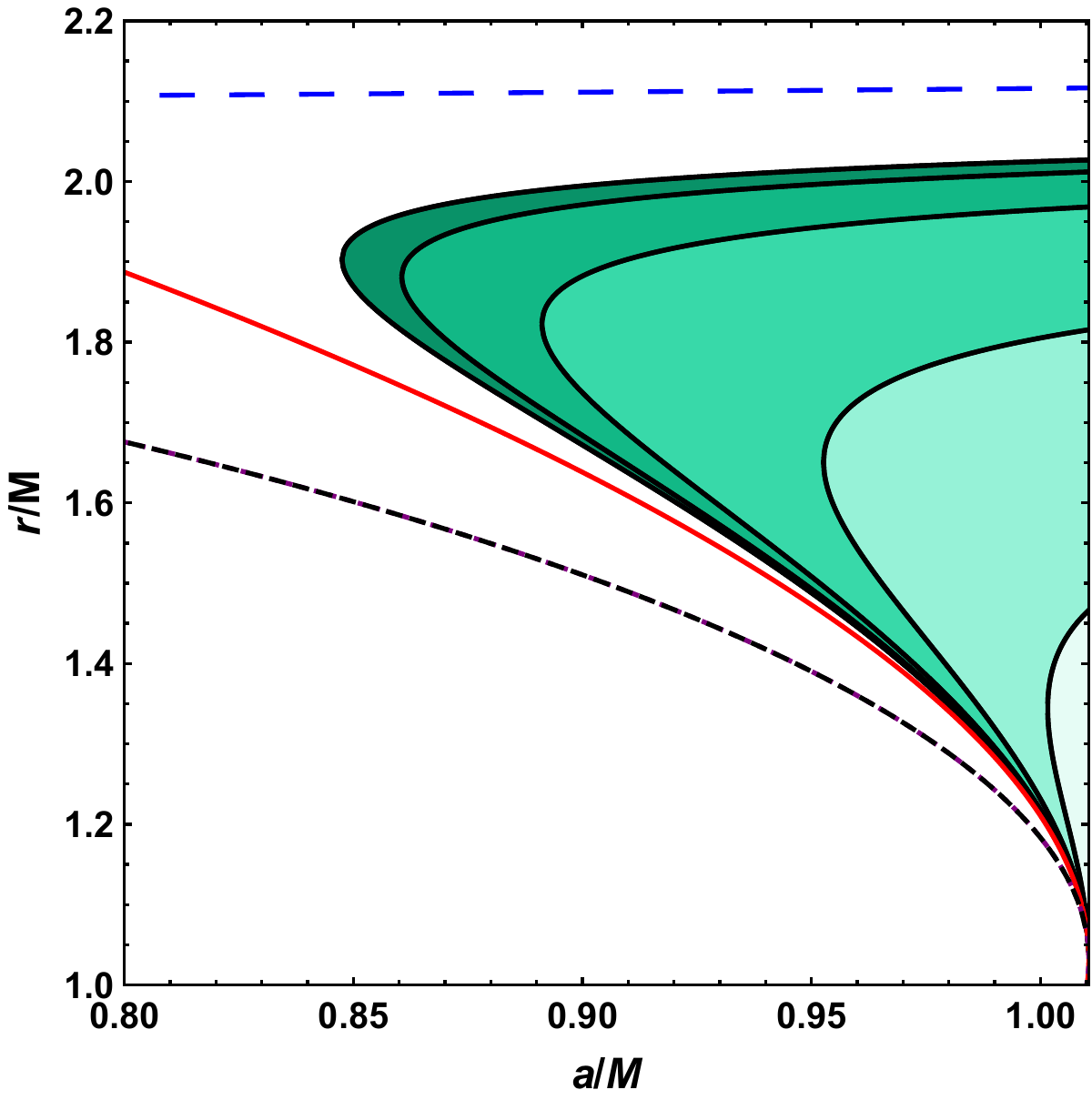}}
\subfigure[$\xi=\pi/12,~\Lambda M^2=0.05$]{\label{ar_5d}
\includegraphics[width=5cm]{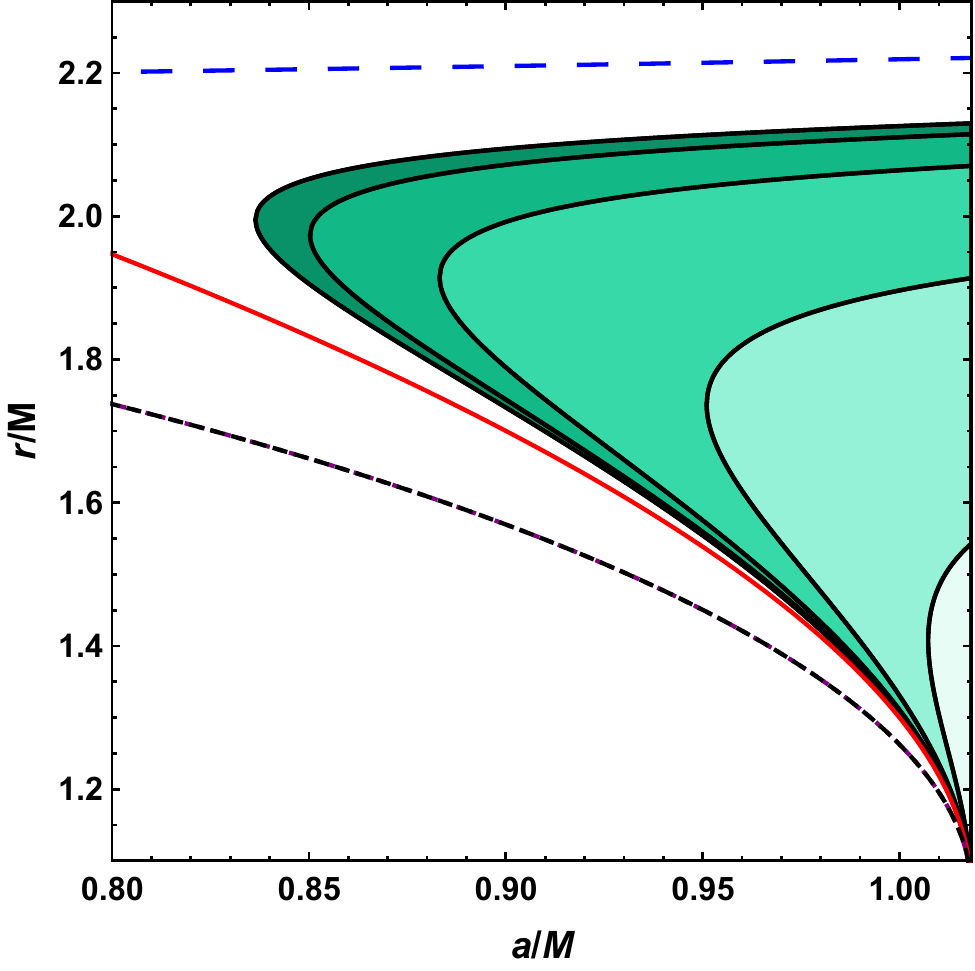}}
\subfigure[$\xi=\pi/12,~\Lambda M^2=0.09$]{\label{ar_5e}
\includegraphics[width=5cm]{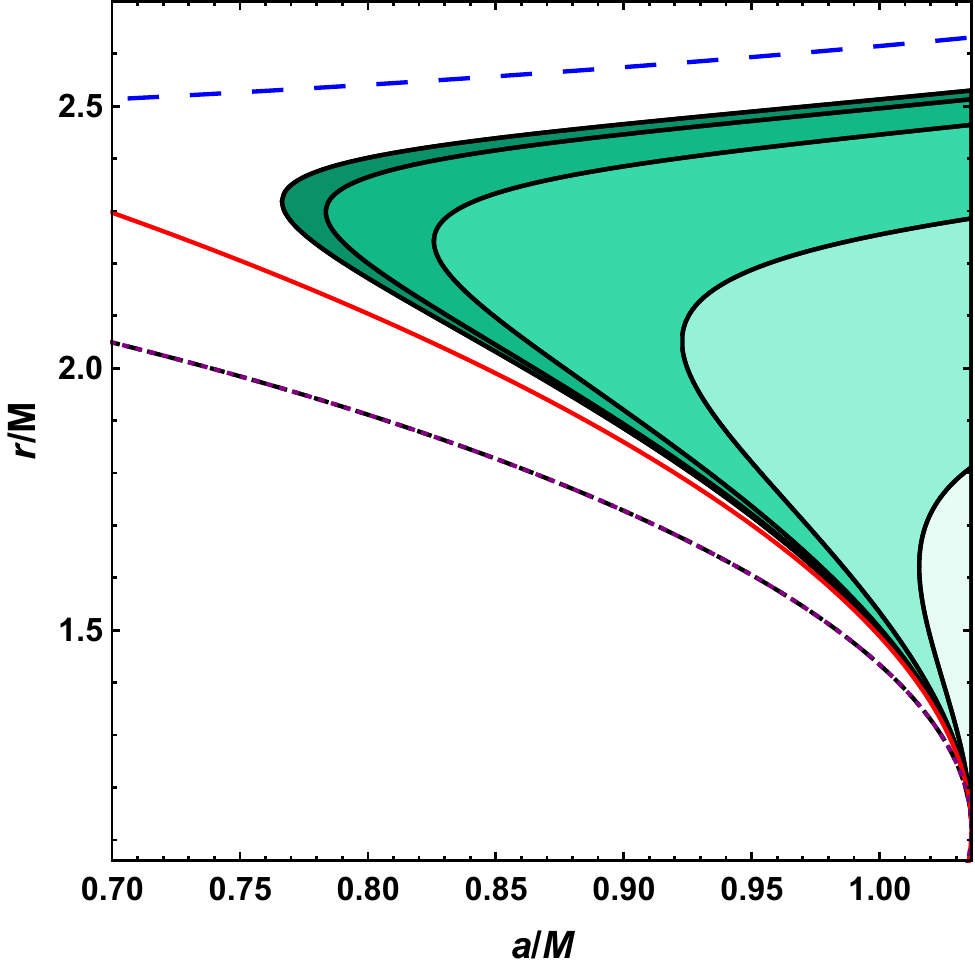}}
\subfigure[$\xi=\pi/12,~\Lambda M^2=0.1$]{\label{ar_5f}
\includegraphics[width=5cm]{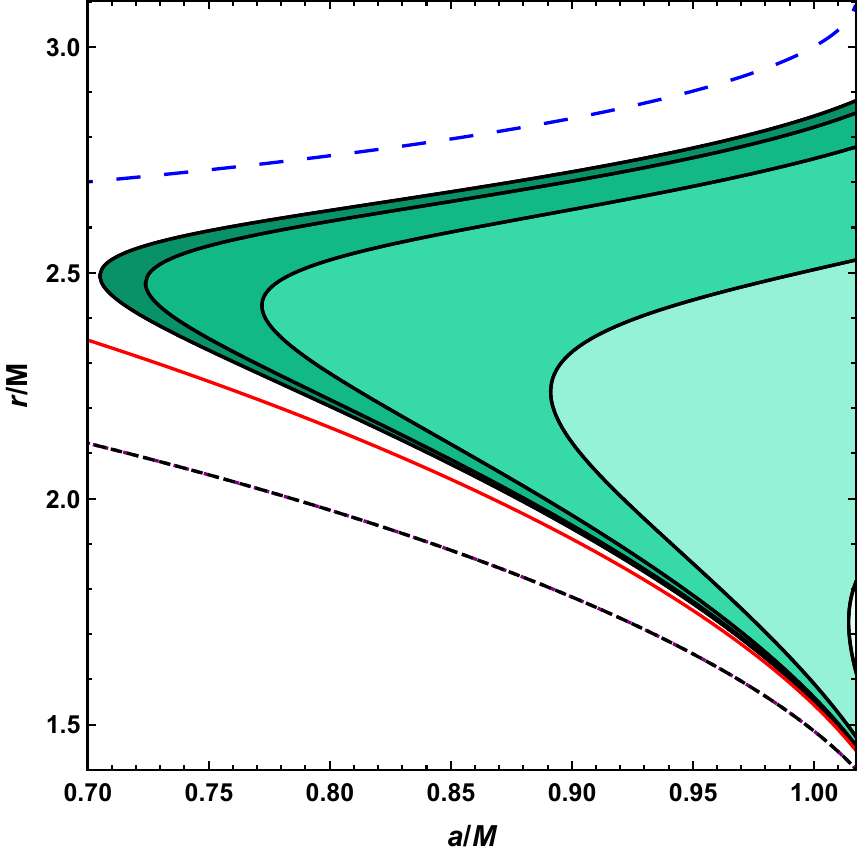}}
}\caption{Regions of the parameter space ($r/M$, $a/M$) for $\epsilon_{+}^{\infty}>0$ and $\epsilon_{-}^{\infty}<0$ with different cosmological constant $\Lambda$. The color regions are for $\epsilon_{-}^{\infty}<0$ with $\xi=\pi/12$ and $\sigma_0$=100, 30, 10, 3, 1 from left to right. Black dot-dashed curves, red solid curves, and blue dashed curves denote the radii of the outer horizon, light ring, and outer ergosphere, respectively. (a) $\Lambda M^2=0$, $a_{max}=1, r_E=2M$. This is the case of Kerr black hole. (b) $\Lambda M^2=0.01$, $a_{max}=1.0033$. (c) $\Lambda=0.03$, $a_{max}=1.0105$. (d) $\Lambda M^2=0.05$, $a_{max}=1.0183$. (e) $\Lambda M^2=0.09$, $a_{max}=1.0360$. (f) $\Lambda M^2=0.1$, $a_{max}=1.0410$.}\label{ar_5}
\end{figure}

\begin{figure}[H]
\center{
\subfigure[$\sigma_0=100,~\Lambda M^2=0$]{\label{ar_6a}
\includegraphics[width=5cm]{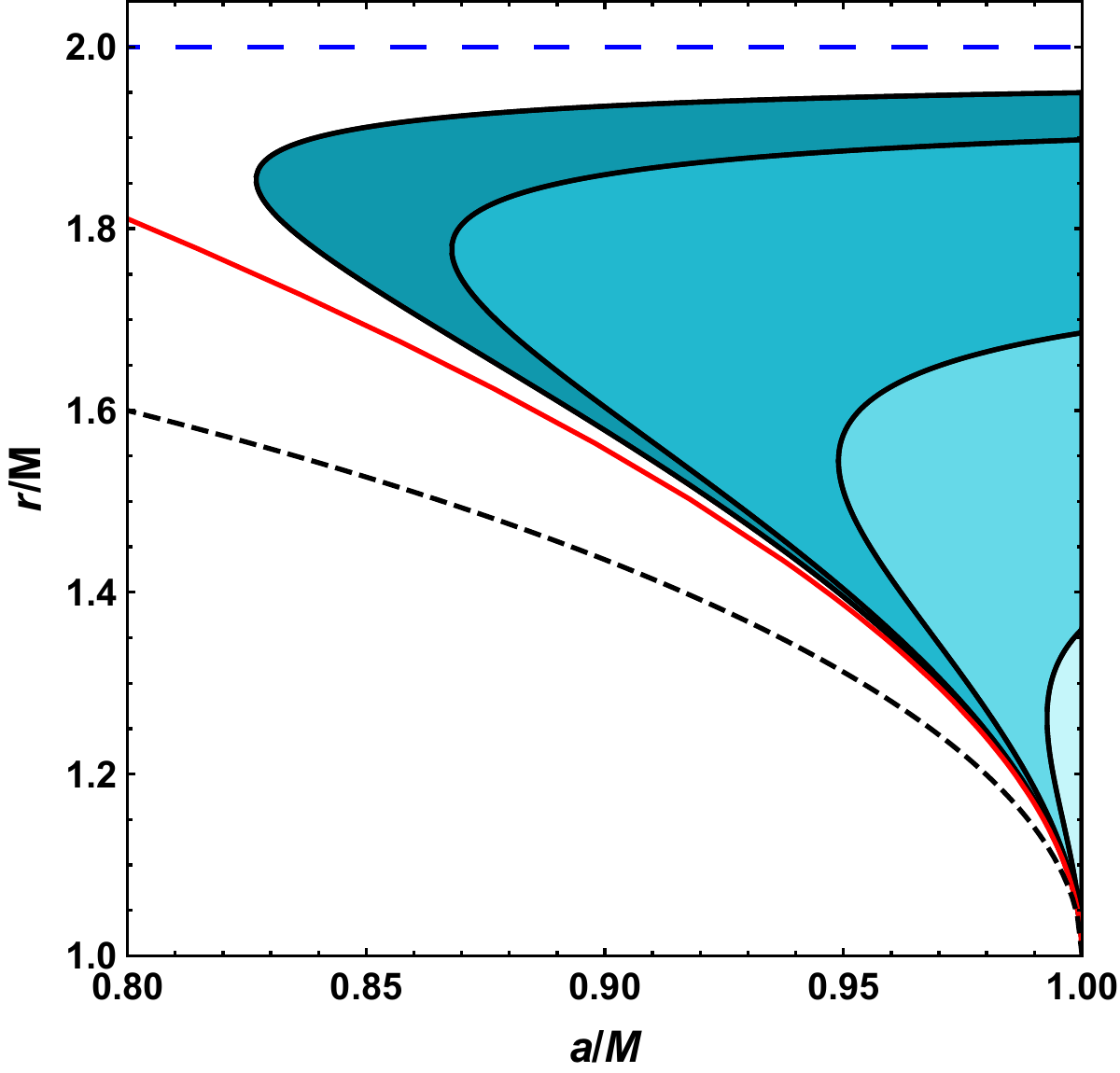}}
\subfigure[$\sigma_0=100,~\Lambda M^2=0.01$]{\label{ar_6b}
\includegraphics[width=5cm]{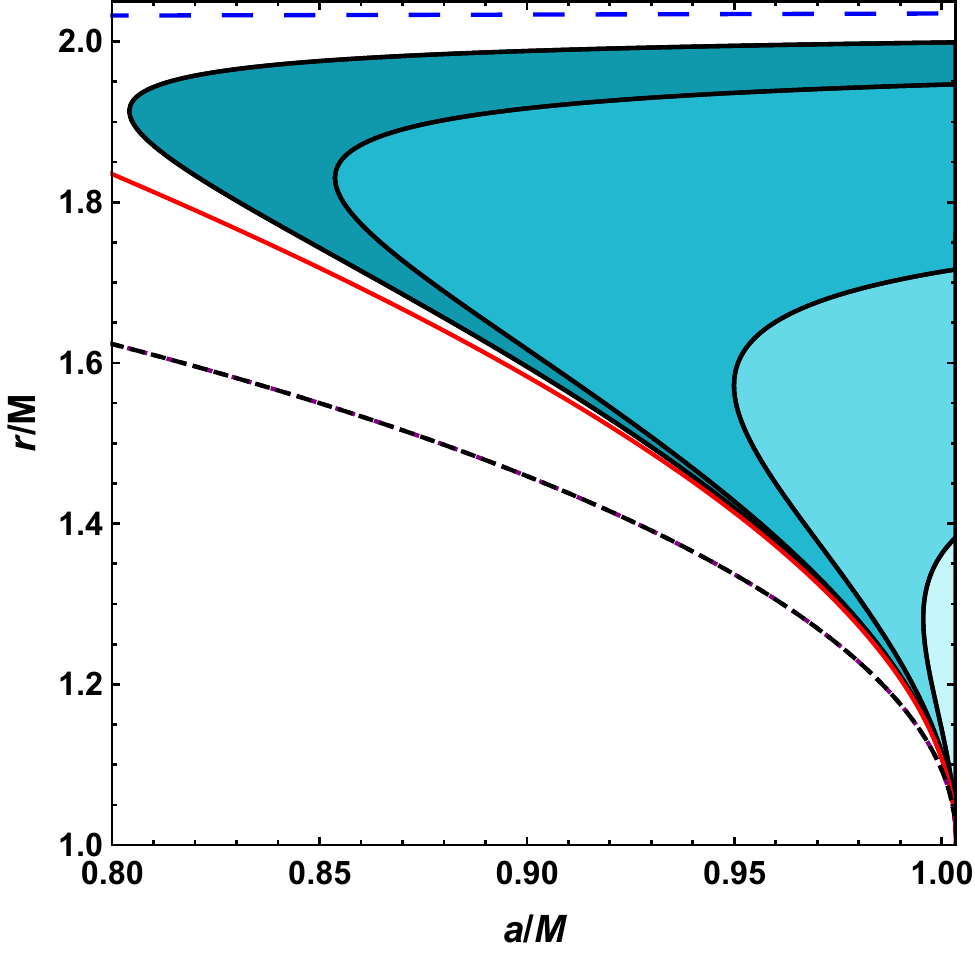}}
\subfigure[$\sigma_0=100,~\Lambda M^2=0.03$]{\label{ar_6c}
\includegraphics[width=5cm]{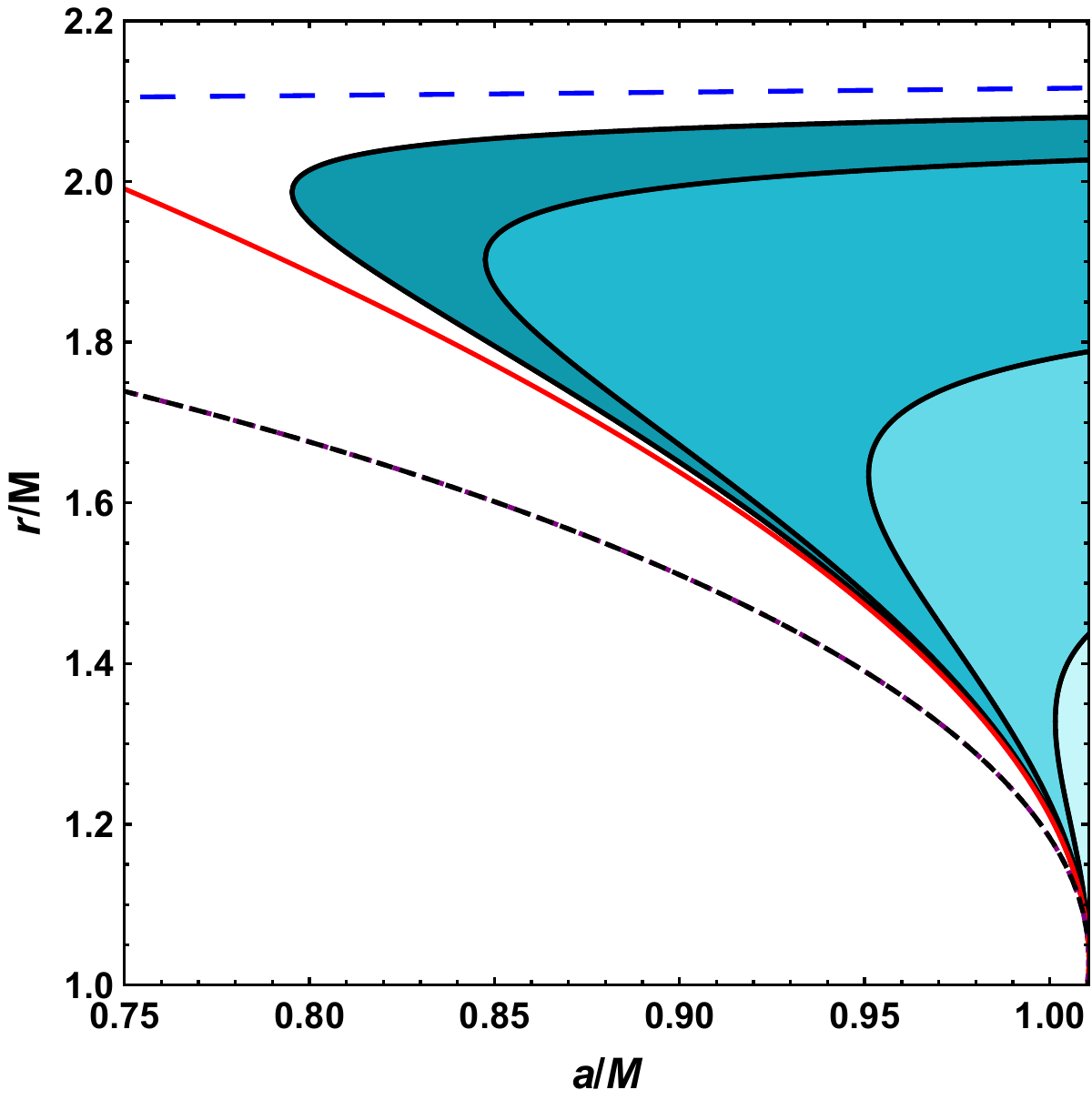}}
\subfigure[$\sigma_0=100,~\Lambda M^2=0.05$]{\label{ar_6d}
\includegraphics[width=5cm]{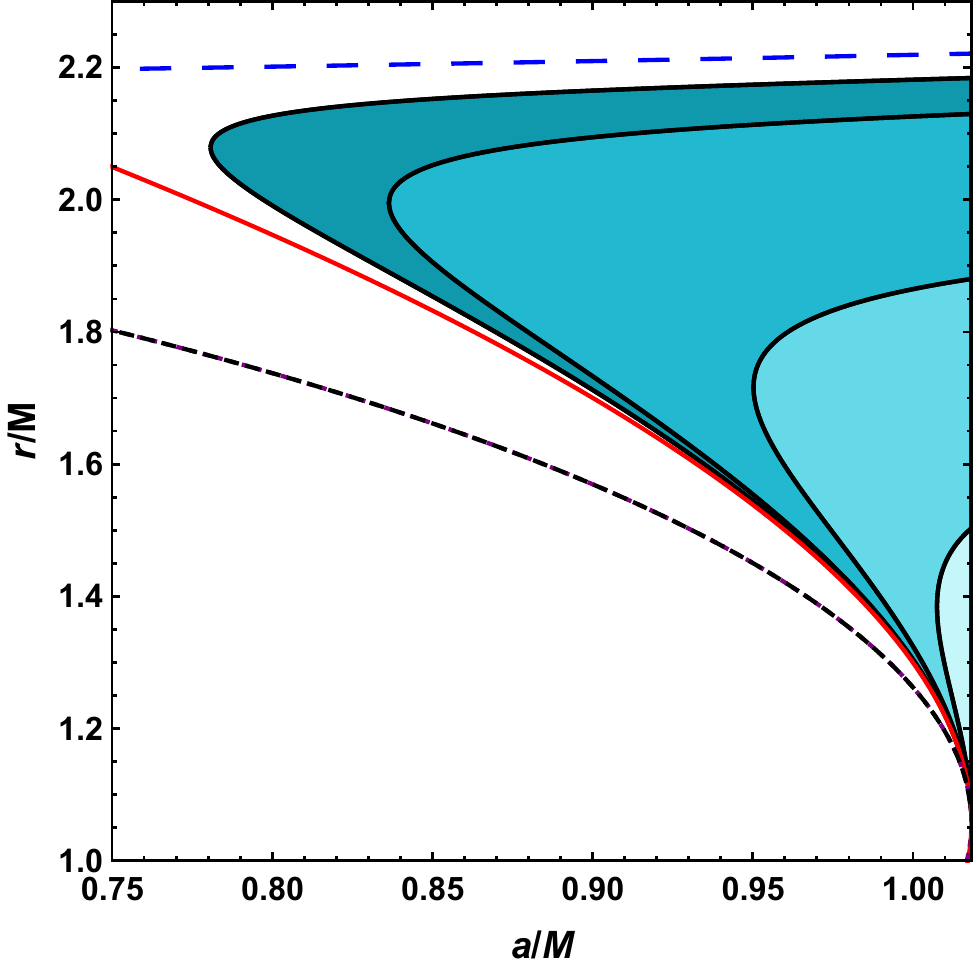}}
\subfigure[$\sigma_0=100,~\Lambda M^2=0.09$]{\label{ar_6e}
\includegraphics[width=5cm]{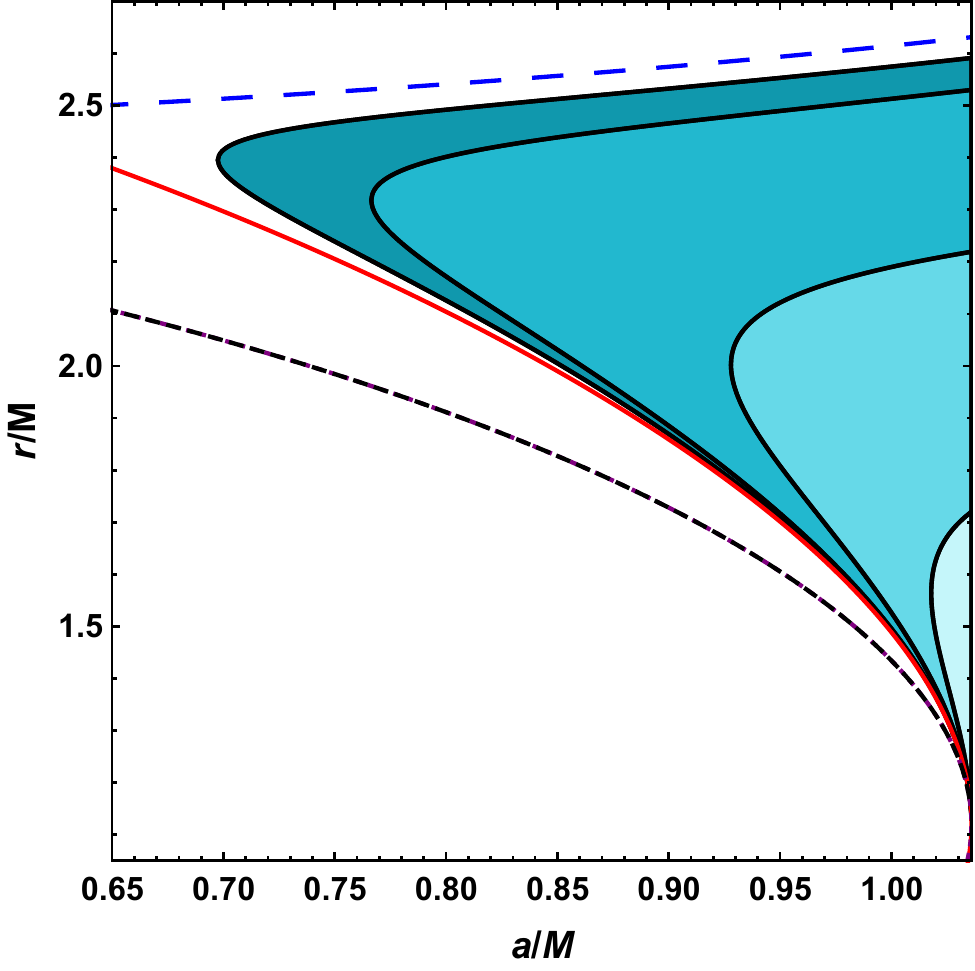}}
\subfigure[$\sigma_0=100,~\Lambda M^2=0.1$]{\label{ar_6f}
\includegraphics[width=5cm]{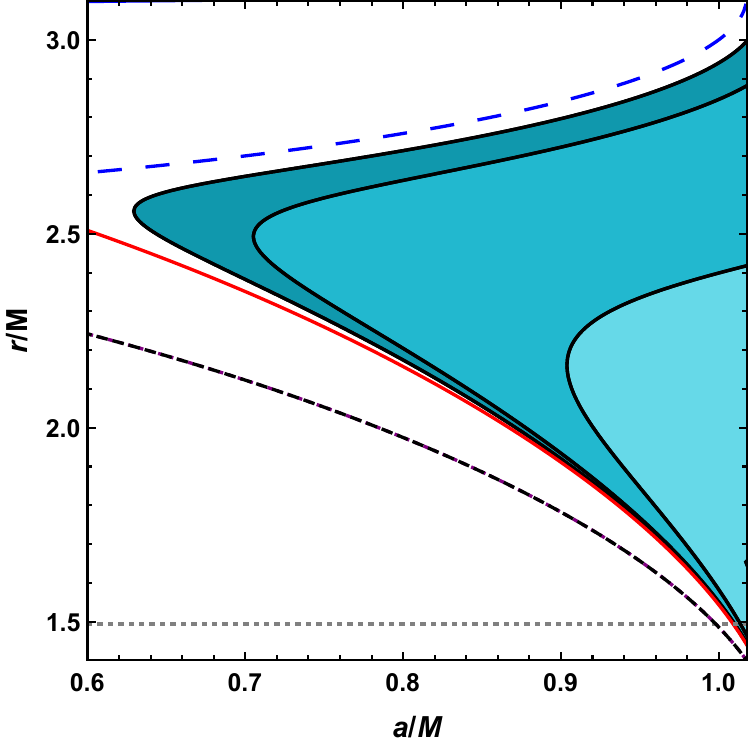}}}
\caption{Regions of the parameter space ($r/M$, $a/M$) for $\epsilon_{+}^{\infty}>0$ and $\epsilon_{-}^{\infty}<0$ with different cosmological constant $\Lambda M^2$. The color regions are for $\epsilon_{-}^{\infty}<0$ with plasma magnetization parameter $\sigma_0$=100 and $\xi$=$\pi/20, \pi/12, \pi/6, \pi/4 $ from left to right. Black dot-dashed curves, red solid curves, and blue dashed curves denote the radii of the outer horizon, light ring, and outer ergosphere, respectively. (a) $\Lambda M^2=0$, $a_{max}=1$ (Kerr black hole case). (b) $\Lambda M^2=0.01$, $a_{max}=1.0033$. (c) $\Lambda M^2=0.03$, $a_{max}=1.0105$. (d) $\Lambda M^2=0.05$, $a_{max}=1.0183$. (e) $\Lambda M^2=0.09$, $a_{max}=1.0360$. (f) $\Lambda M^2=0.1$, $a_{max}=1.0410$.}\label{ar_6}
\end{figure}

Next, we turn our attention to the $a/M-r/M$ parameter space with the results displayed in Figs. \ref{ar_5} and \ref{ar_6}. In Fig. \ref{ar_5}, we take a small orientation angle $\xi=\pi/12$ while varying the cosmological constant $\Lambda M^2$=0, 0.01, 0.03, 0.05, 0.09, and 0.1, respectively. For each subfigure, we show the negative $\epsilon_-^{\infty}$ in shaded regions for the magnetization parameter $\sigma_0$=1, 3, 10, 30, and 100. Similarly, all these shaded regions fall in the ergosphere region as expected. More significantly, the circular photon orbits shown in red solid curves provide stricter lower bounds for these regions. In Fig. \ref{ar_5a}, it just gives the result for the Kerr black holes with vanishing $\Lambda$. By increasing the magnetization parameter $\sigma_0$, the shaded regions allowed to extract black hole rotational energy via the magnetic reconnection extend to larger values of $r/M$ and lower values of black hole spin $a/M$. This result is exactly consistent with that given in Ref. \cite{Comisso:2020ykg}. For nonzero $\Lambda$, such pattern also holds. For a fixed $\sigma_0$, there exists a minimum value of black hole spin $a/M$. With the increase of $\Lambda$, this value slightly decreases. Furthermore, the ergosphere expands with $\Lambda$, leading to the fact that the shaded region could extend to a larger value of $r/M$. This phenomenon is obvious for the rapidly spinning black holes with $\Lambda M^2$=0.1, see Fig. \ref{ar_5f}.

As mentioned above, the larger magnetization parameter leads to the larger region to implement the energy extraction. So taking $\sigma_0=100$, we show these regions satisfying the condition (\ref{condition}) in $a/M-r/M$ parameter space in Fig. \ref{ar_6}. For each figure, we set the orientation angle $\xi=$ $\pi/4,~\pi/6,~\pi/12$, and $\pi/20$ from right to left. It is easy to find that all these cases with different values of $\Lambda M^2$ are similar. Also, they are quite similar to these cases shown in Fig. \ref{ar_5}. Comparing with the Kerr black hole case given in Fig. \ref{ar_6a} , we observe that the cusp of the shaded regions with $\xi=\pi/20$ can even approach $a/M=0.64$ (Fig. \ref{ar_6f}) smaller than 0.83 of the Kerr black hole.

In conclusion, we uncover two potential aspects of the cosmological constant $\Lambda$ on the energy extraction via the magnetic reconnection mechanism by comparing with the Kerr black holes. This first one is that the energy extraction becomes possible in a slowly spinning Kerr-dS black hole rather than its Kerr counterpart. Another one is that the maximum value of the black hole spin increases with the cosmological constant, which provides us a new chance to extract black hole rotational energy from a rapidly spinning black hole with spin $a/M>1$.

\section{Power and efficiency via magnetic reconnection}
\label{paevmr}

In this section, we turn to evaluate the power and efficiency of black hole energy extraction through the magnetic reconnection from a Kerr-dS black hole. The extracted net energy is mainly dependent on the amount of the plasma particles with negative energy at infinity produced by the magnetic reconnection and absorbed by the black hole. This indicates that the energy extraction rate is induced by the corresponding reconnection rate. As a result, the energy extracted power from the black hole by the escaping plasma, $P_{\text{extr}}$, can be evaluated as
\begin{eqnarray}\label{eqofpextr}
 P_{\text{extr}}=-\epsilon_{-}^{\infty}w_0A_{\text{in}}U_{\text{in}}.
\end{eqnarray}
In the collisionless and collisional regimes, $U_{\text{in}}$=$\mathcal{O}(10^{-1})$ and $\mathcal{O}(10^{-2})$ \cite{Comisso:2016ima}. Quantity $A_{in}$ denotes the cross-sectional area of the inflowing plasma, which can be well estimated as $A_{in}\sim (r_E^2-r_{po}^2)$.

\begin{figure}[H]
\center{
\subfigure[$\xi=\pi/12,~a/M=0.999,~\sigma_0=100$]{\label{pextrrlambda_7a}
\includegraphics[width=5.7cm]{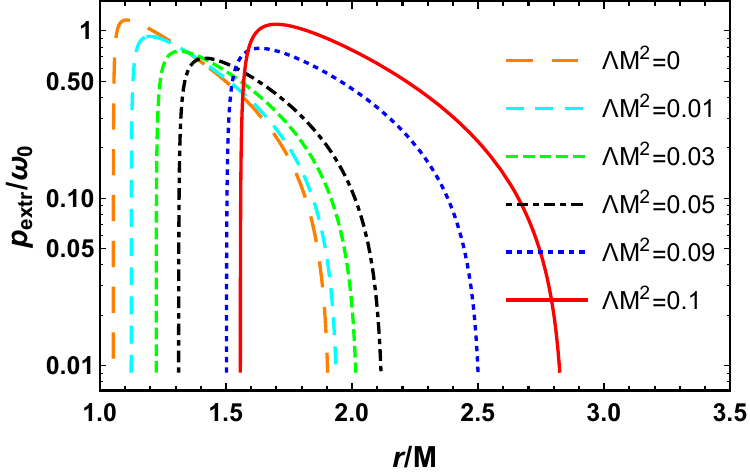}}
\subfigure[$\xi=\pi/12,~\Lambda M^2=0.01,~\sigma_0=100$]{\label{pextrra_7b}
\includegraphics[width=5.7cm]{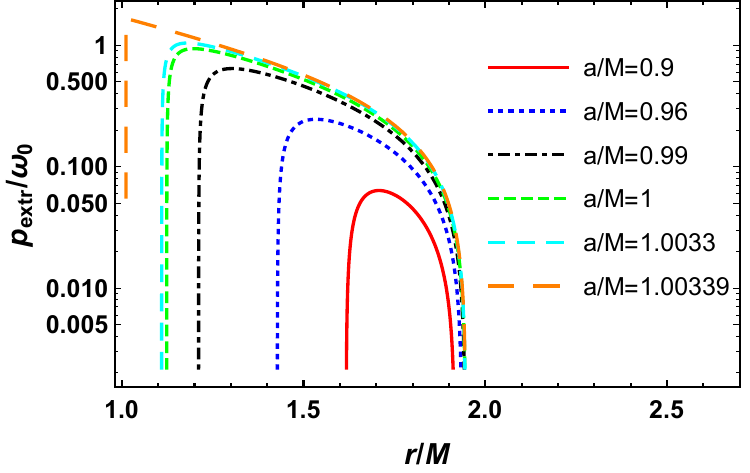}}
\subfigure[$\xi=\pi/12,~a/M=0.999,~\sigma_0=100$]{\label{pextrlambdar_7c}
\includegraphics[width=5.7cm]{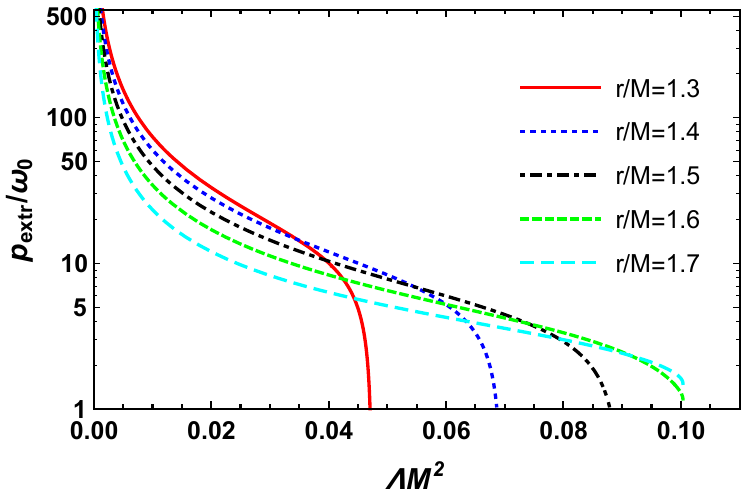}}
}
\caption{Log plot of the power $P_{\text{extr}}/w_0$ as a function of the dominant $X$-point location $r/M$ and cosmological constant $\Lambda M^2$ for a rapidly spinning black hole. } \label{ppPowers_7}
\end{figure}

The behaviors of the power $p_{extr/\omega_0}$ for rapidly spinning black holes with $\xi=\pi/12$, $\sigma_0=100$, and $U_{\text{in}}$=0.1 is shown in Fig. \ref{ppPowers_7}.  In Fig. \ref{pextrrlambda_7a}, we plot the power as a function of $r/M$ of the $X$-point with the black hole spin $a/M$=0.999. The cosmological constant $\Lambda M^2$=0, 0.01, 0.03, 0.05, 0.09, and 0.1 from left to right. For each case, there exists a peak near the limiting circular photon orbit, and then the power decreases with $r/M$. We can also find that the maximal power slightly decreases with $\Lambda$, which indicates the cosmological constant reduces the ability of the energy extraction. However, the possible location of the $X$-point is shifted towards larger $r/M$. Thus the magnetic reconnection could occur far away from the black hole with a larger $\Lambda$. On the other hand, for the fixed $\Lambda M^2$=0.01, see Fig. \ref{pextrra_7b}, the power increases with the black hole spin, which implies that the rapidly spinning black hole leads to high power, exactly the same result given in \cite{Comisso:2020ykg}. In order to uncover the effect of the cosmological constant on the power, we show it as a function of $\Lambda M^2$ for different locations of the $X$-point in Fig. \ref{pextrlambdar_7c}. For each $r/M$, the power shares a similar pattern. It drops rapidly at first, then slows down, and finally it drops rapidly around some certain values of $\Lambda M^2$. Although the power of the energy extraction will be reduced by the cosmological constant, the rapidly spinning Kerr-dS black hole with $a/M>1$ can provide a potential energy source for the magnetic reconnection process. It is also noteworthy that the energy can be extracted at a large $r/M$ of $X$-point for nonvanishing cosmological constant.

In this process of the energy extraction, magnetic field energy is a key, which can operate such mechanism. Furthermore, magnetic energy also plays the role to redistribute the angular momentum of the particles such that those particles with energy $\epsilon_{+}^{\infty}$ escape to infinity, and those with energy $\epsilon_{-}^{\infty}$ fall into the black hole. If $\epsilon_{-}^{\infty}$ is negative, then more energy will be carried out by these escaping particles. As a result, one can define the efficiency of the plasma energization process via magnetic reconnection as \cite{Comisso:2020ykg}
\begin{eqnarray}
 \eta=\frac{\epsilon_{+}^{\infty}}{\epsilon_{+}^{\infty}+\epsilon_{-}^{\infty}}.
\end{eqnarray}
Obviously, the rotational energy will be extracted when $\eta>1$.

\begin{figure}[H]
\center{
\subfigure[$\xi=\pi/12,~\Lambda M^2=0.01,~\sigma_0=100$]{\label{etara_8a}
\includegraphics[width=7cm]{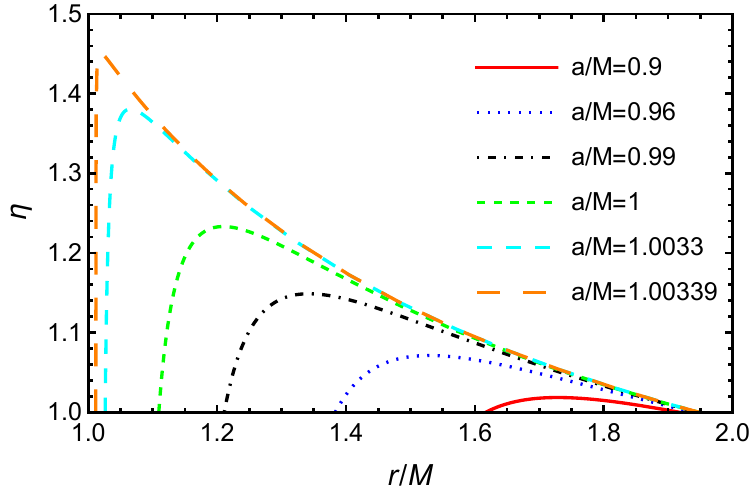}}
\subfigure[$\xi=\pi/12,~a/M=0.999,~\sigma_0=100$]{\label{etarlambda_8b}
\includegraphics[width=7cm]{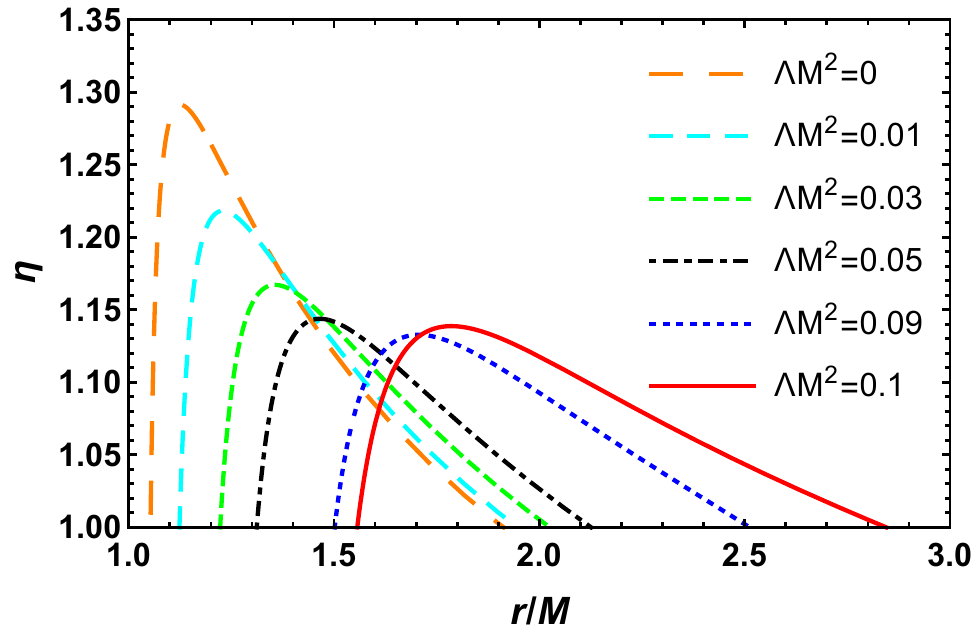}}}
\caption{ Efficiency $\eta$ of the reconnection process for a rapidly spinning black hole. (a) $\eta$ vs $r/M$ with $\Lambda M^2=0.01$ and $\xi=\pi/12$. (b) $\eta$ vs $r/M$ with $a/M=0.999$ and $\xi=\pi/12$. } \label{etar_8}
\end{figure}

In Fig. \ref{etar_8}, we plot the efficiency $\eta$ as a function of the $X$-point location with $\xi=\pi/12$ and $\sigma_0=100$. When $\Lambda M^2=0.01$, we exhibit the efficiency $\eta>1$ in Fig. \ref{etara_8a} for black hole spin $a/M$=0.9$\sim$1.00339. For small spin with $a/M=0.9$, we can see that the range of the $X$-point location is bounded in $r/M\in$ (1.6, 2.0) for $\eta>1$ to extract black hole energy. Meanwhile, the peak of the efficiency is slight larger than 1. With the increase of the black hole spin, the lower bound extends to small $r/M$. The peak also increases. With further increase in the black hole spin such that $a/M$=1, the peak can approach $\eta=1.2$. Since the maximum value of the Kerr-dS black hole spin is beyond 1, we see that the maximum efficiency can be further increased. In particular, when the black hole spin $a/M$ approaches 1.00339, $\eta$ can exceed 1.45. On the other hand, taking $a/M$=0.999, we show the efficiency in Fig. \ref{etarlambda_8b} for different values of $\Lambda M^2$. The behavior is quite similar to that of Fig. \ref{etara_8a}. Although the cosmological constant reduces the peak of the efficiency, it widens the range of the location of the $X$-point. For example, for the Kerr black hole, the possible $X$-point has a width $\Delta r/M\approx$1, while it becomes near 1.5 for the Kerr-dS black hole with $\Lambda M^2$=0.1. So comparing with the Kerr black holes, Kerr-dS black holes have an advantage in a larger radial distance to achieve an efficient energy extraction process.

\begin{figure}[H]
\center{
\subfigure[$\xi=\pi/12,~a/M=0.999,~\sigma_0=100$]{\label{etalambdar_9a}
\includegraphics[width=7cm]{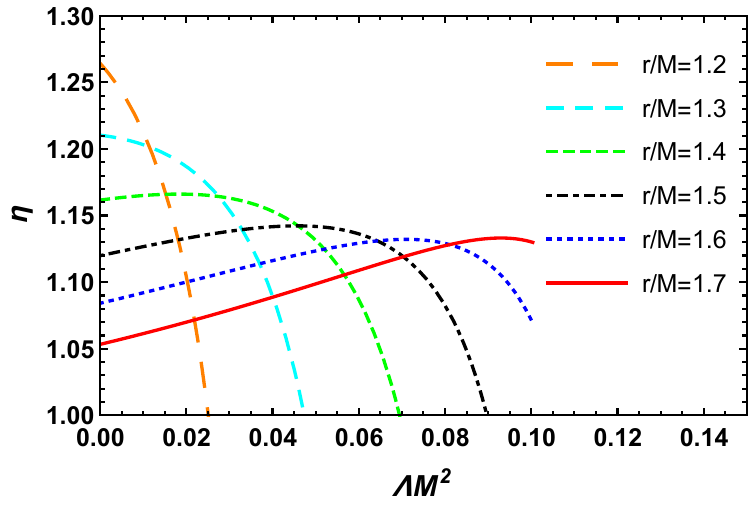}}
\subfigure[$\xi=\pi/12,~\Lambda M^2=0.01,~\sigma_0=100$]{\label{etaar_9b}
\includegraphics[width=7cm]{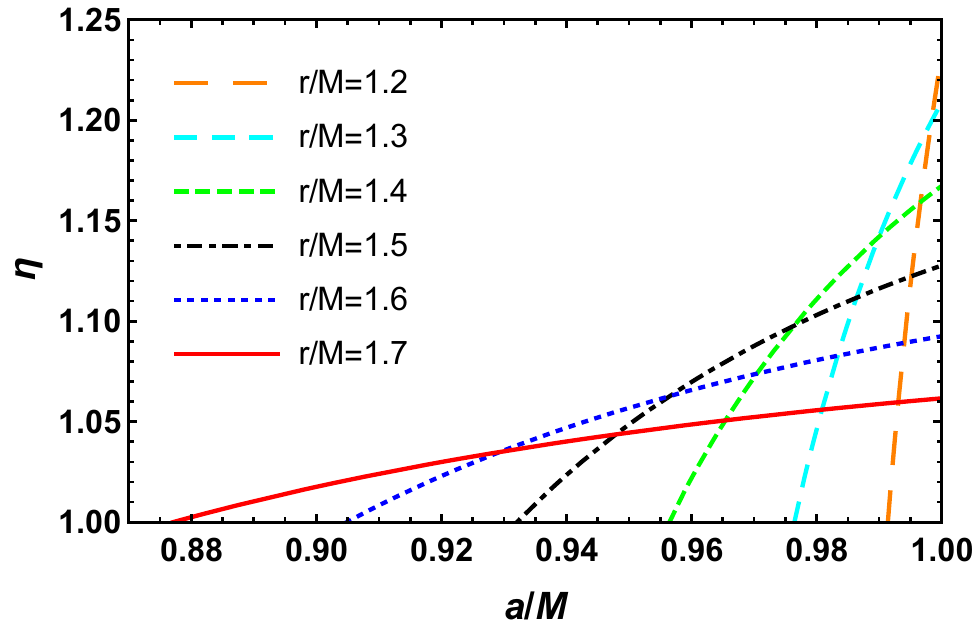}}}
\caption{Efficiency $\eta$ of the reconnection process for a rapidly spinning black hole. (a) $\eta$ vs $\Lambda M^2$ with $a/M=0.999$, and $\xi=\pi/12$. Note that we have truncated the cosmological constant at $\Lambda M^2$=0.10035. (b) $\eta$ vs $a/M$ with $\Lambda M^2=0.01$ and $\xi=\pi/12$.} \label{etaaL_9}
\end{figure}

Moreover, we wonder about the behaviors of the efficiency $\eta$ as function of $\Lambda$ or $a$ for fixed $X$-point location. The results are clearly exhibited in Fig. \ref{etaaL_9} with the fixed $X$-point at $r/M$=1.2, 1.3, 1.4, 1.5, 1.6, and 1.7. From Fig. \ref{etalambdar_9a}, we find that, for small $r/M$, the Kerr black hole indeed admits a higher efficiency than the Kerr-dS black hole. However, when the location of $r/M(>1.4)$ is far away from the black hole, the result reverses. A peak appears at some certain values of $\Lambda M^2$, indicating that the Kerr-dS black hole has advantages to extract the black hole energy if these magnetic reconnections happen at large $r/M$. Note that we have truncated the cosmological constant at $\Lambda M^2$=0.10035, beyond which such energy extraction process is impossible, see Fig. \ref{parameterspace}. In addition, for small cosmological constant $\Lambda M^2$=0.01, we observe in Fig. \ref{etaar_9b} that the high efficiency can be achieved for high spin and small $r/M$.

\begin{figure}[H]
\center{
\subfigure[$\xi=\pi/12,~a/M=0.999,~\sigma_0=100$]{\label{pkvspkds_11a}
\includegraphics[width=7cm]{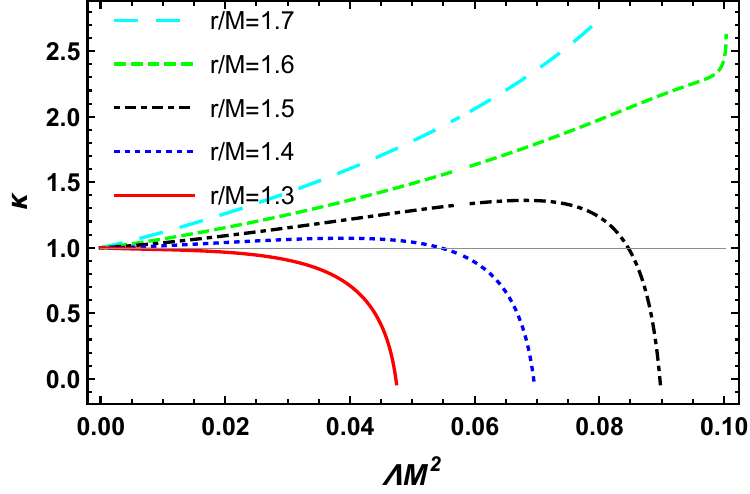}}
\subfigure[$\xi=\pi/12,~\sigma_0=100,~r/M=1.6$]{\label{pkvspkds_11b}
\includegraphics[width=7cm]{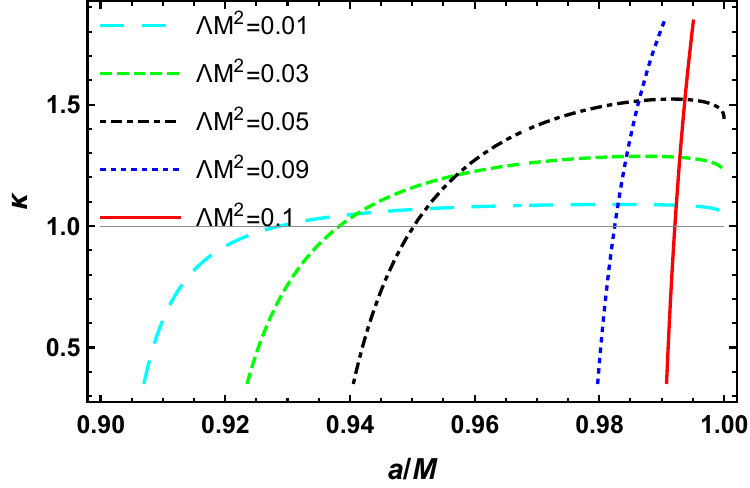}}
}
\caption{Power ratio $\kappa$ for a rapidly spinning black hole. (a) $\kappa$ as a function of $\Lambda M^2$ with $\xi=\pi/12$, $a/M=0.999$, and $\sigma_0=100$ for different values of $r/M$. (b) $\kappa$ as a function of $a/M$ with $\xi=\pi/12$, $a/M=0.999$, and $r/M=1.6$ for different values of $\Lambda M^2$.} \label{pkerr1}
\end{figure}

Finally, we would like to examine whether the magnetic reconnection mechanism in Kerr-dS black holes is more efficient than the original Comisso-Asenjo mechanism in Kerr black holes. For this purpose, we define a ratio of the power
\begin{eqnarray}
 \kappa=\frac{P_{extr}^{Kerr-dS}}{P_{extr}^{Kerr}}.
\end{eqnarray}
The behavior of the power ratio $\kappa$ is exhibited in Fig. \ref{pkerr1}. By fixing the orientation angle of the particle and the magnetization of the plasma, we plot the ratio $\kappa$ against the cosmological constant in Fig. \ref{pkvspkds_11a} for the rapidly spinning black hole with $a/M=0.999$. For the small cosmological constant, we observe that energy extraction from the Kerr and Kerr-dS black holes is comparable regardless of $r/M$. While with the increase of $\Lambda M^2$, the case becomes interesting. For $r/M=1.3$, one can see that $\kappa$ is smaller than 1, which implies that the energy extraction from the Kerr black hole is more efficient than the Kerr-dS black hole. However, when $r/M$=1.4 or 1.5, the Kerr-dS black hole has a high power for small $\Lambda M^2$, while the Kerr black hole dominates at large $\Lambda M^2$. If $r/M$ is large enough, the Kerr-dS black hole always has a high power for arbitrary cosmological constant. Taking $\xi=\pi/12$, $\sigma_0=100$, and $r/M$=1.6, we plot the ratio $\kappa$ as a function of black hole spin $a/M$ in Fig. \ref{pkvspkds_11b}. For different values of $\Lambda M^2$, ratio $\kappa$ behaves similarly. With the increase of black hole spin $a/M$, it increases from a certain value below 1 and approaches a maximal value, and finally gets a slight decrease when the maximal black hole spin is reached. This indicates that the Kerr black hole has an advantage in the low spin cases. Instead, the Kerr-dS black hole is more promising in the high spin cases to extract the energy.

\begin{figure}[htp]
\center{\includegraphics[width=7cm]{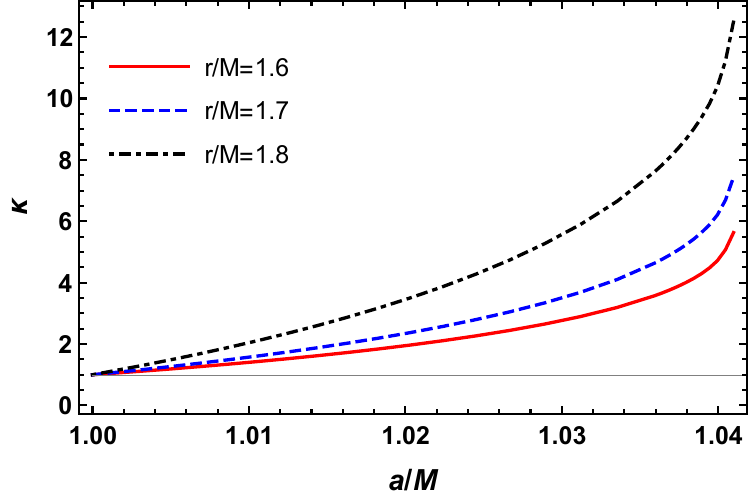}}
\caption{The ratio $\kappa$ with both Kerr-dS and Kerr black holes being extremal black holes with $\sigma_0$=100 and $\xi=\frac{\pi}{12}$.} \label{pevspkerrrM_12}
\end{figure}

As we have shown, the presence of the cosmological constant admits a higher power than the Kerr black holes. On the other hand, the bound of the maximal black hole spin also increases with the cosmological constant, such that the black hole spin can be beyond the Kerr bound $a/M$=1. Here we show the power ratio $\kappa$ of the extremal Kerr-dS and Kerr black holes in Fig. \ref{pevspkerrrM_12}. For the maximal $a/M$=1, the extremal Kerr-dS black hole agrees with the extremal Kerr black hole, and thus we have $\kappa$=1 as expected. Moreover, with the increase in the maximal spin bound, we observe that the ratio $\kappa$ is monotonically increasing. Such ratio can amount to 12 when $r/M$=1.8. This result indicates that the increase of black hole spin bound by the cosmological constant may be a main factor to approach a higher power of energy extraction.

\section{Conclusions and discussions}
\label{Conclusion}

In this paper, we have studied the energy extraction via the magnetic reconnection mechanism for the Kerr-dS black holes. The effects of the cosmological constant on this process were investigated in detail.

Compared with the Kerr black hole, we found that there are two significant aspects for the cosmological constant to influence this energy extraction mechanism. The first one is that the positive cosmological constant will increase the spin bound $a/M=1$ of the Kerr black hole, see Fig. \ref{parameterspace}. Another one is the ergosphere. For the Kerr black hole, the radius of the ergosphere in the equatorial plane is $r_{E}=2M$ independent of the black hole spin. However for the Kerr-dS black hole, it depends on both the cosmological constant and the black hole spin, and it is larger than the Kerr case. This result will lead to a relatively wider region for the $X$-points.

At first, several characteristic radii for the Kerr-dS black holes were studied. For example, the radii of the black hole horizon, the ergosphere, and the circular photon orbit increase with $\Lambda M^2$. This allows us to consider the energy extraction in the appropriate region for the $X$-point. A detailed study uncovers that this region widens with $\Lambda M^2$, indicating that the Kerr-dS black hole can lead to a more efficient energy extraction mechanism.

Then we gave a brief introduction for the energy at infinity per enthalpy $\epsilon_{+}^{\infty}$ and $\epsilon_{-}^{\infty}$. In order to extract the black hole energy, it requires $\epsilon_{+}^{\infty}>0$ and $\epsilon_{-}^{\infty}<0$. For this purpose, we examined the parameter regions that can implement the mechanism. Respectively varying the cosmological constant $\Lambda$ and location of $X$-point, we found that the Kerr-dS black hole is better able to implement the energy extraction than the Kerr black hole for larger values of $\Lambda$ and $X$-point.

The power and efficiency of the energy extraction via the magnetic reconnection mechanism were also calculated. It is found that, although the presence of the cosmological constant results the decrease of the maximum power $P_{extr}/\omega_0$ and efficiency, it will dominate at larger value of $r/M$ for the $X$-point. Especially, the maximum value of the efficiency will go beyond that of the Kerr counterpart for the Kerr-dS black hole with $a/M>1$.

We also calculated the ratio of the power for the Kerr-dS and Kerr black holes through the original Comisso-Asenjo mechanism. The results clearly reveal that the ratio is larger than 1 at large $X$-point location and rapidly spinning black hole spin with $a/M>1$.

In conclusion, in this paper, we studied the potential possibility of extracting the black hole rotational energy via the magnetic reconnection mechanism when the positive cosmological constant is included. Our studies show that the Kerr-dS black hole indeed has advantages when the black hole spin $a/M>1$ and the dominant reconnection $X$-point is far away from the event horizon. These results disclose the potential effects of the cosmological constant on the energy extraction via the magnetic reconnection. It is also expected to extend this study to other spinning hairy black holes or traversable wormholes.

\section*{Acknowledgements}
This work was supported by the National Natural Science Foundation of China (Grants No. 1207510, No. 12047501 and No. 11965016), and Qinghai Science and Technology Plan, No. 2020-ZJ-728.

\end{document}